\documentclass[namedreferences]{solarphysics}
\usepackage[optionalrh]{spr-sola-addons}
\usepackage{graphicx}
\usepackage{url}             

\usepackage{solaheader}
\newcommand{\solareceiveddate}{15 December 2011}
\newcommand{\solaaccepteddate}{27 July 2012}
\newcommand{\doi}{012-0095-5}

\renewcommand{\TODAY}{12 September 2012}

\newcommand{\B}{\mbox{\boldmath$B$}}

\newcommand{\curl}{\mbox{\boldmath$\nabla$}\mathbf{\times}}

\newcommand{\wrat}{w}
\newcommand{\wind}{q}
\newcommand{\qmod}{\wind_{_m}} 
\newcommand{\wmod}{\wrat_{_m}} 
\newcommand{\qmodamp}{\wind_{_c}} 
\newcommand{\qobsamp}{\wind_{_{c,{\rm obs}}}} 
\newcommand{\qbamp}{Q}
\newcommand{\basis}{h}
\newcommand{\myie}{{\it i.e.}}
\newcommand{\myeg}{{\it e.g.}}
\newcommand{\myl}{l}
\newcommand{\twist}{T} 

\newcommand{\nvart}{n_a}
\newcommand{\bvart}{B_a}

\newcommand{\rw}{\rho}

\newcommand{\tw}{\phi} 
\newcommand{\pw}{\tau}
\newcommand{\xw}{z}
\newcommand{\yw}{x}
\newcommand{\zw}{y}

\newcommand{\rl}{R_t}
\newcommand{\al}{a_t}
\newcommand{\ql}{q_t}
\newcommand{\bt}{B_t}
\newcommand{\rc}{\sigma}
\newcommand{\mysc}{s}
\newcommand{\fgl}{L}

\newcommand{\aap}{    {\it Astron. Astrophys.}}

\newcommand{\apj}{    {\it Astrophys. J.}}
\newcommand{\apjl}{    {\it Astrophys. J. Lett.}}

\newcommand{\apss}{   {\it Astrophys. Space Sci.}}

\newcommand{\gafd}{   {\it Geophys. Astrophys. Fluid Dyn.}}
\newcommand{\grl}{    {\it Geophys. Res. Lett.}}

\newcommand{\jastp}{  {\it J. Atmos. Solar Terr. Phys.}}

\newcommand{\jfm}{    {\it J. Fluid Mech.}}

\newcommand{\jgra}{    {\it J. Geophys. Res. (Space Phys.) }}
\newcommand{\mnras}{  {\it Mon. Not. Roy. Astron. Soc.}}

\newcommand{\rspsa}{  {\it Roy. Soc. London Proc. Ser. A}}

\newcommand{\solphys}{{\it Solar Phys.}}

\begin{document}
\begin{article}
\begin{opening}

\title{Least-Squares Fitting Methods for Estimating the Winding Rate in Twisted Magnetic-Flux Tubes}

\author{A.D.~\surname{Crouch}}

\runningauthor{A.D.~Crouch}
\runningtitle{Fitting Methods for Winding-Rate Estimation}

\institute{A.D.~Crouch\\NorthWest Research Associates, 3380 Mitchell Lane, Boulder, CO 80301, USA\\email: \url{ash@cora.nwra.com}}


\begin{abstract}
We investigate least-squares fitting methods for estimating the winding rate of field lines about the axis of twisted magnetic-flux tubes.
These methods estimate the winding rate by finding the values for a set of parameters that correspond to the minimum of the discrepancy between vector magnetic-field measurements and predictions from a twisted flux-tube model.
For the flux-tube model used in the fitting, we assume that the magnetic field is static, axisymmetric, and does not vary in the vertical direction.
Using error-free, synthetic vector magnetic-field data constructed with models for twisted magnetic-flux tubes, we test the efficacy of fitting methods at recovering the true winding rate.
Furthermore, we demonstrate how assumptions built into the flux-tube models used for the fitting influence the accuracy of the winding-rate estimates.
We identify the radial variation of the winding rate within the flux tube as one assumption that can have a significant impact on the winding-rate estimates.
We show that the errors caused by making a fixed, incorrect assumption about the radial variation of the winding rate can be largely avoided by fitting directly for the radial variation of the winding rate.
Other assumptions that we investigate include the lack of variation of the field in the azimuthal and vertical directions in the magnetic-flux tube model used for the fitting, and the inclination, curvature, and location of the flux-tube axis. 
When the observed magnetic field deviates substantially from the flux-tube model used for the fitting, we find that the winding-rate estimates can be unreliable.
We conclude that the magnetic-flux tube models used in this investigation are probably too simple to yield reliable estimates for the winding rate of the field lines in solar magnetic structures in general, unless additional information is available to justify the choice of flux-tube model used for the fitting.
\end{abstract}


\end{opening}

\section{Introduction}

Methods for reliably estimating the twist present in solar magnetic-flux concentrations are required for several areas of solar-physics research. Some examples include:
The latitudinal variation of the twist present in active regions (\myeg{} \opencite{alpha_lat}; \opencite{sigma98}; \opencite{2001ApJ...549L.261P}; \opencite{2006JGRA..11112S01N}) provides constraints for models of the generation and dynamics of magnetic-flux tubes in the solar convection zone (\myeg{} \opencite{1994GeoRL..21..241R}; \opencite{lintonetal96}; \opencite{1996ApJ...458..380L}; \opencite{1996ApJ...464..999L}; \opencite{sigma98}; \opencite{1998ApJ...493..480F}; \opencite{1999mhsl.conf...75G}; \opencite{2000ApJ...540..548A}; \opencite{2004ApJ...615L..57C}).
The twist present in a magnetic-flux tube can provide information about the susceptibility of the flux tube to the kink instability, where magnetic twist is rapidly converted to writhe (deformation of the flux-tube axis), which may play a role in eruptive events such as coronal mass ejections (\myeg{} \opencite{1972SoPh...22..425R}; \opencite{HoodPriest1979}; \opencite{1990Ap+SS.166..289C}; \opencite{Vellietal1990}; \opencite{lintonetal96}; \opencite{1998ApJ...494..840L}; \opencite{vanderLindenHood1998}, \citeyear{vanderLindenHood1999}; \opencite{1998A+A...333..313B}; \opencite{1997SoPh..172..249B}, \citeyear{Baty2001}; \opencite{fan03}, \citeyear{fan04}; \opencite{lea03}; \opencite{tor03}; \opencite{tor04}; \opencite{kink}; \opencite{Nandyetal2008}).
Estimates for the twist present in photospheric magnetic-flux concentrations are useful inputs for coronal magnetic-field simulations (\myeg{} \opencite{2006ApJ...641..577M}, \citeyear{2006ApJ...642.1193M}; \opencite{Nandyetal2008}; \opencite{2008SoPh..247..103Y}; \opencite{2009ApJ...699.1024Y}; \opencite{2010ApJ...709.1238Y}) and provide constraints for simulations of emerging twisted magnetic-flux tubes (\myeg{} \opencite{1998ApJ...492..804E}; \opencite{1998MNRAS.298..433H}; \opencite{fan03}, \citeyear{fan04}; \opencite{2008A+A...479..567M}; \opencite{2009A+A...503..999H}; \opencite{2009A+A...507..995M}; \opencite{2010ApJ...720..233C}; \opencite{2011ApJ...735..126T}).
The magnetic twist is closely related to the twist component of the magnetic helicity (\myeg{} \opencite{BergerField1984} \opencite{1992RSPSA.439..411M}).

In a simple magnetic-flux tube, the magnetic twist can be decomposed into the product of two quantities: the length of the flux tube [$\myl$], and the winding rate [$\wind$] (\myie{} the angle that a field line rotates about the flux-tube axis per unit length along the axis), assuming that the winding rate does not vary in the direction along the flux-tube axis.
The focus of this article is methods for estimating the winding rate of the field lines in magnetic-flux tubes from vector magnetic-field measurements at a single observation height.
Several such methods currently exist that we now briefly review.

The $\alpha_{\rm best}$ method (\myeg{} \opencite{alpha_lat}; \opencite{sigma98}; \opencite{alpha2}; \opencite{alpha1};  \opencite{2001ApJ...549L.261P};  \opencite{lea03}; \opencite{2004ApJ...611.1149H}; \opencite{kink}; \opencite{2006JGRA..11112S01N}) computes a single (constant) value of $\alpha$, 
the force-free parameter in the equation $\curl \B =  \alpha \B$, 
that provides the least-squares best-fit between a linear force-free magnetic-field model (computed using the observed vertical component of the magnetic field) and the observed horizontal components of the magnetic field, usually for an entire active region.
This method was used by \inlinecite{alpha_lat},  \inlinecite{sigma98}, \inlinecite{2001ApJ...549L.261P}, \inlinecite{2004ApJ...611.1149H}, and \inlinecite{2006JGRA..11112S01N} to show how  $\alpha_{\rm best}$ in active regions depends on latitude.
By averaging over the entire active region this technique is less susceptible to noise than methods that compute $\alpha$ by differentiating the magnetic field (discussed later).
However, the large-scale averaging means that the magnetic twist is not probed on spatial scales smaller than an entire active region (\myeg{}  \opencite{alpha1}; \opencite{alpha2}; \opencite{kink}).
Another limitation of the $\alpha_{\rm best}$ method is that the magnetic field at the solar photosphere, where the vector fields are typically measured, is not force-free (\myeg{} \opencite{1995ApJ...439..474M}).

The $\alpha_{\rm best}$ method was also used by \inlinecite{lea03} to compute the winding rate and, subsequently, the twist present in active regions associated with eruptive events, in order to determine whether the kink instability was the mechanism responsible for the eruption.
To determine the winding rate, \citeauthor{lea03}  assumed $\alpha_{\rm best} =  2 \wind$ (see also \opencite{2006JGRA..11112S01N}); a similar assumption is invoked by \inlinecite{2009ApJ...700..199T} and \inlinecite{2009ApJ...702L.133T} but for a method different from the $\alpha_{\rm best}$ method.
\inlinecite{kink} pointed out that this relationship between the force-free parameter [$\alpha$] and the winding rate [$\wind$] is generally only true at the axis of an axisymmetric magnetic-flux tube, for a flux tube that is not thin with an axis that is straight and vertical (\myeg{} \opencite{1989GApFD..48..217F}).
Furthermore, \inlinecite{kink}  showed that the $\alpha_{\rm best}$ method tends to underestimate the  winding rate on the small spatial scales that may be relevant to determining whether a magnetic-flux system in an active region is susceptible to the kink instability (see also, \myeg, \opencite{alpha1}; \opencite{alpha2}).

\inlinecite{kink} introduced the $\alpha_{\rm peak}$ method for estimating the winding rate of the magnetic-field lines in the vicinity of the axis of a twisted magnetic-flux tube. 
The $\alpha_{\rm peak}$ method assumes that the axis corresponds to the region of largest $\alpha_z$ for a flux tube with a uniform winding rate.
The relationship $\alpha_z = 2 \wind$ is then used to estimate the winding rate at the flux-tube axis.
As shown in Table~1 of \citeauthor{kink} the relative discrepancy between the inferred and true winding rates at the axis recovered with this method is roughly 10\,\%, when tested with synthetic data from the simulation results of \citeauthor{fan03}~(\citeyear{fan03}, \citeyear{fan04}); note that $\wind_{\rm peak} = \alpha_{\rm peak}/2 = - 10$~radians $\fgl^{-1}$ for the simulation results of \citeauthor{fan03}~(\citeyear{fan03}, \citeyear{fan04}), where $\fgl$ is a length scale.
Two aspects of this method make the results potentially sensitive to noise in the measurements.
First, this method samples only a handful of points in the vicinity of the inferred location of the flux-tube axis.
Second,  $\alpha_z$ is computed by differentiating the vector magnetic-field measurements (see also, \myeg, \opencite{1994ApJ...425L.117P}; \opencite{alpha1}; \opencite{alpha2}; \opencite{2003ApJ...597L..73N}; \opencite{2005ApJ...629.1135H}; \opencite{2009ApJ...697L.103S}; \opencite{2009ApJ...700..199T}; \opencite{2009ApJ...702L.133T}), and differentiation is generally a noise amplifying operation (unless techniques are used  to suppress the influence of noise).

\inlinecite{Nandyetal2008} introduced a least-squares fitting technique for estimating the winding rate in magnetic-flux tubes from vector magnetic-field data.
To estimate the winding rate in a magnetic-flux tube, the fitting method outlined by \citeauthor{Nandyetal2008} finds the values of a set of model parameters, related to the winding rate, that correspond to the minimum of the discrepancy between the observations and the predictions from a prescribed flux-tube model (hereafter, the {\it fitting model}).
This type of approach has several potential advantages over methods that use the force-free parameter. 
These include:
i) fitting methods do not assume that the magnetic field is force-free;
ii) fitting methods do not require derivatives of the magnetic-field measurements and are, therefore, likely to be less sensitive to noise than methods that do involve such derivatives;
iii) fitting methods use many observation points made within a solar magnetic-flux tube and are, therefore,  likely to be less sensitive to noise than methods that only use a few points; and
iv) fitting methods may be able to probe the winding rate throughout the flux-tube interior, not only in the vicinity of the axis.
However, there are still several possible sources of error in the approach outlined by \citeauthor{Nandyetal2008} that need to be investigated.
For example, \citeauthor{Nandyetal2008} assumed that the magnetic field in the fitting model is axisymmetric and does not vary in the vertical direction. 
Moreover, \citeauthor{Nandyetal2008} assumed that the winding rate inside the flux tube varies with radius [$r$] according to $\wind (r) = \wind_{\rm fit} + c/r$, where the best fit value of $\wind_{\rm fit}$ is their estimate for the winding rate and $c$ is another model parameter; this equation implies that the winding rate is infinite at $r=0$ for $c \neq 0$.

The purpose of this investigation is to demonstrate how assumptions  built into models for twisted magnetic-flux tubes can affect estimates for the winding rate inferred with least-squares fitting methods.
For this purpose we rely on error-free synthetic vector magnetic-field measurements for which the correct winding rate is known.
The outline of this article is as follows:
The details of the fitting model and the assumptions built into it are described in Section~\ref{sec_mod}.
The details of the fitting method and the metrics used to test its performance are described in Section~\ref{sec_fit}.
In Section~\ref{sec_sing} we examine a version of the fitting method that makes a fixed assumption about the radial variation of the winding rate (in a similar fashion to \opencite{Nandyetal2008}).
We test this fitting method with synthetic data constructed with models for twisted magnetic-flux tubes that have a radially-dependent winding rate that differs from the fitting model.
We find that the inferred winding rate is sensitive to the assumed radial variation of the winding rate built into the fitting model.
Subsequently, in Section~\ref{sec_multi} we present and test a fitting method that infers  the radial variation of the winding rate.
We find that this method can generally perform better than fitting methods that have a fixed radial variation of the winding rate, provided that the magnetic field in the fitting model and that used to construct the synthetic data is static, axisymmetric, and does not vary in the vertical direction.
In Section~\ref{sec_assump} and the Appendix, we determine the sensitivity of the winding-rate estimates retrieved by fitting methods to the other assumptions built into the fitting model, such as the neglect of vertical and azimuthal variations in the magnetic field.
We also apply the fitting method to synthetic vector magnetic-field measurements generated with a twisted toroidal magnetic-flux tube model, which is substantially different from the fitting model. 
In this case, we find that the winding-rate estimates can be inaccurate. Consequently, we do not apply the fitting method to real vector magnetic-field data.
In Section~\ref{sec_conc} we draw conclusions.

\section{The Fitting Model}
\label{sec_mod}

The fitting model is a simplified model for an isolated, static, twisted magnetic-flux tube.
We assume that error-free, ambiguity-resolved, vector magnetic-field measurements $(B_x, B_y, B_z)$ of a magnetic-flux tube are available at a set of discrete locations on an $x$\,--\,$y$-plane at constant height $z$, taken to approximate a layer of constant optical depth within a limited field of view.
To describe the fitting model we work in a cylindrical coordinate system $(r, \theta, z)$, assuming that the flux-tube axis is straight, parallel to the $z$-direction, and located at $r=0$, where $r$ is the radial coordinate and $\theta$ is the azimuthal angle measured counterclockwise from the positive $x$-axis to the  position vector in the  $x$\,--\,$y$-plane.
We assume that the magnetic field in the fitting model does not vary in the $\theta$- or $z$-directions.
Under these assumptions, the radial component of the field in the fitting model must be zero for the magnetic field to be divergence-free and for the radial component of the field to be finite at the flux-tube axis.
For this class of flux tubes, the winding rate of a magnetic-field line at a given position
(\myie{} the angle that a field line rotates about the flux-tube axis per unit length along the axis) 
is given by 
$\qmod (r)  = B_\theta(r) / ( r B_z(r) ) $, 
where 
$B_\theta = - \sin \theta \, B_x + \cos \theta \, B_y$
is the azimuthal component of the field (\myeg{} \opencite{1984smh..book.....P}).
In these flux-tube models, the winding rate is independent of $\theta$ and $z$, but can vary in the radial direction.

\begin{figure}[ht]
\begin{center}
\includegraphics[width=0.49\textwidth]{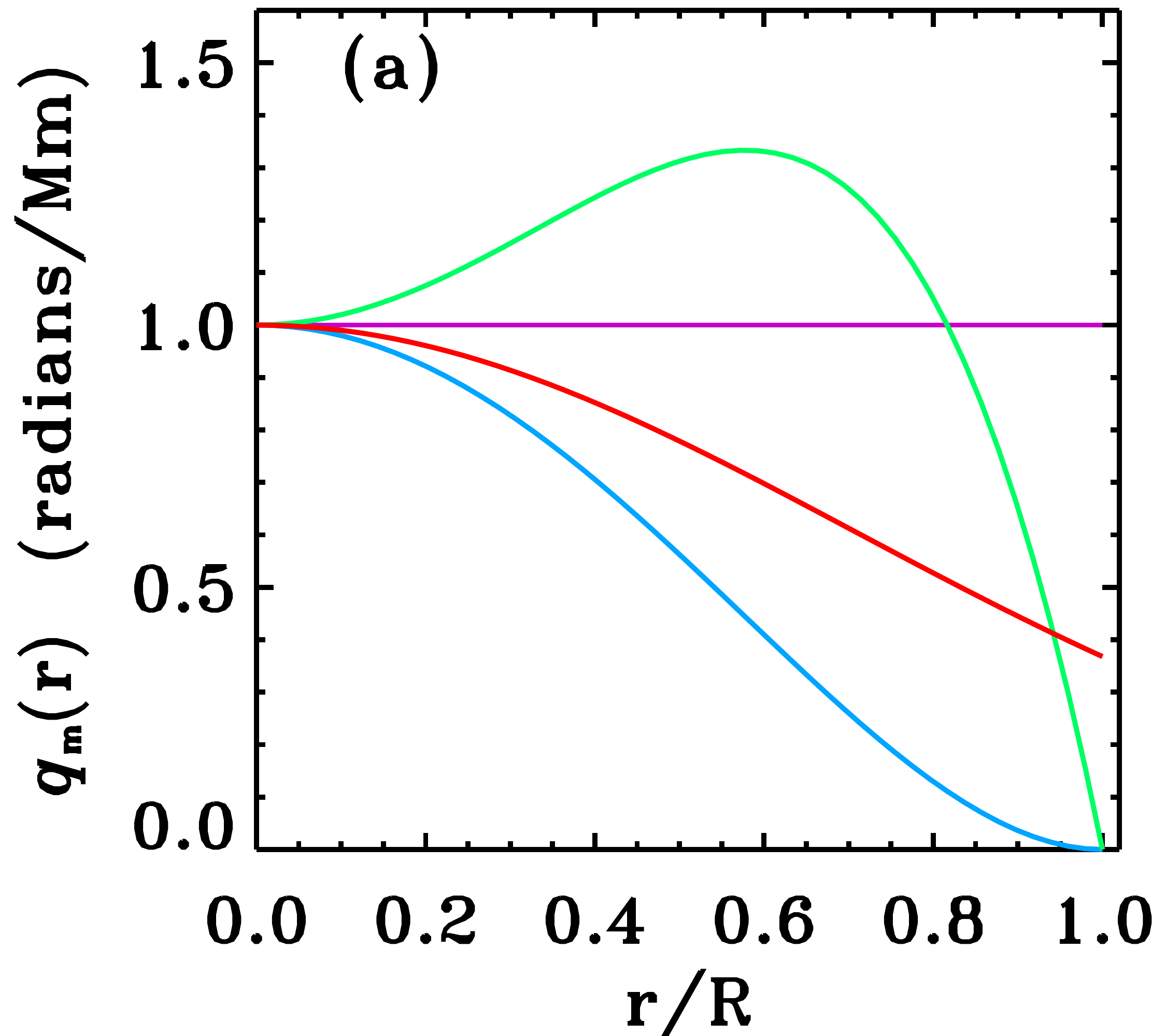}\hfil\includegraphics[width=0.49\textwidth]{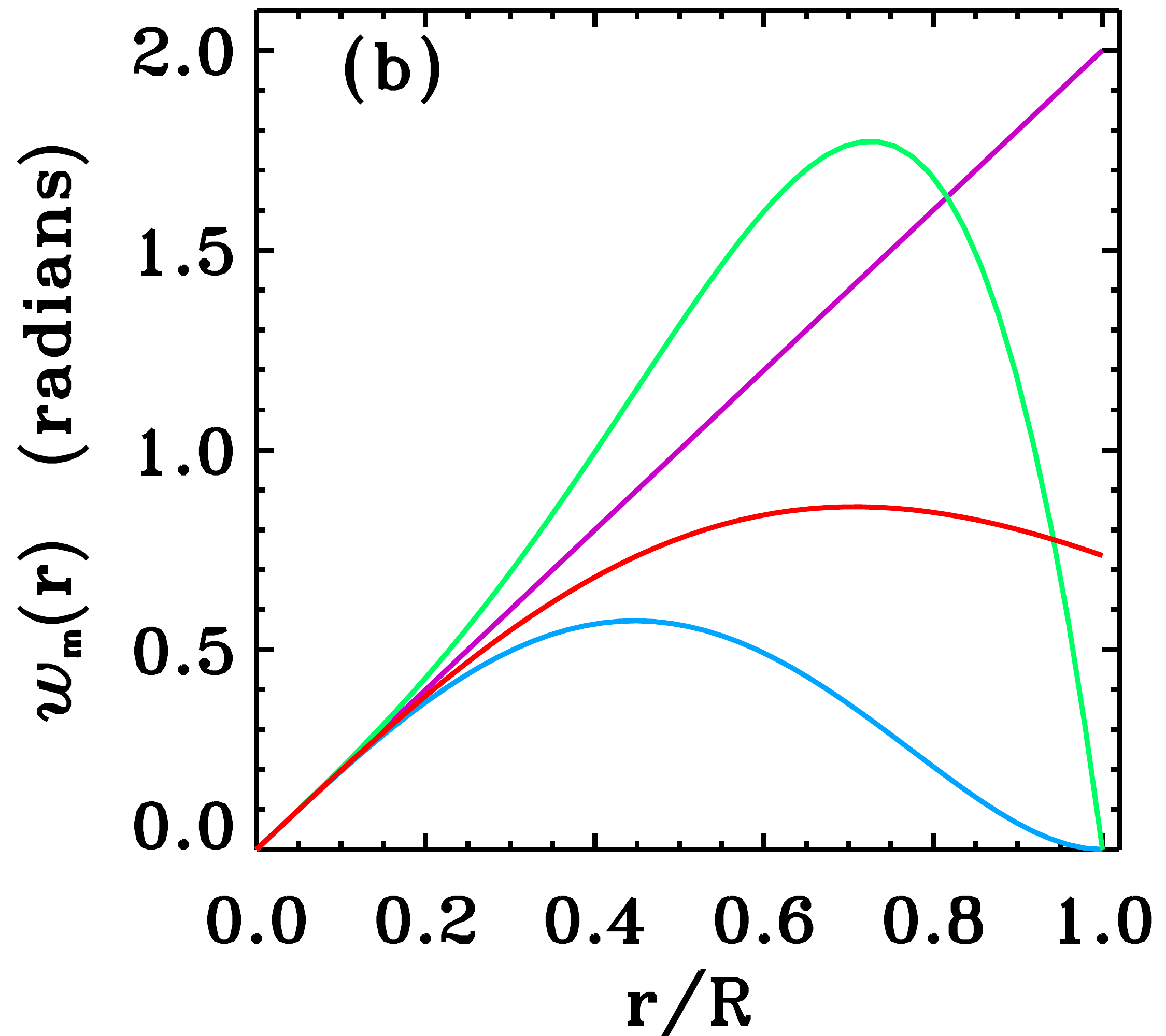}
\end{center}
\caption{(a) Winding rate [$\qmod = B_\theta / ( r B_z ) $] as a function of dimensionless distance from the flux-tube axis [$r/R$] where $R$ is the radius of the flux tube.
The purple curve is for a flux tube with a uniform winding rate, $q_m(r)=\qmodamp$.
The blue curve is for a flux tube with a winding rate that decreases monotonically to zero at the flux-tube boundary, $q_m(r) = \qmodamp (1-(r/R)^2)^2$.
The red curve is for a flux tube with a winding rate that also decreases monotonically, but does not vanish at the boundary, $q_m(r) = \qmodamp \exp (-(r/R)^2)$.
The green curve is for a flux tube with a winding rate that is non-monotonic and vanishes at the boundary, $q_m(r)=\qmodamp (1 + 2 (r/R)^2 - 3 (r/R)^4)$.
In all cases, $\qmodamp =1$~radian~Mm$^{-1}$ and, thus, the winding rate at the flux-tube axis is $q_m(0)=1$~radian~Mm$^{-1}$.
(b) Same as (a) except for the corresponding ratio, $\wmod = B_\theta / B_z$, as a function of $r/R$ for a flux tube with a radius of $R=2$~Mm.
}
\label{egqw}
\end{figure}

To understand the radial variation of the magnetic field and winding rate permitted in the fitting model we refer to the results of \inlinecite{1989GApFD..48..217F}.
We assume that the magnetic field within the fitting model is finite and continuous, but may have a current sheet at the external boundary of the flux tube.
At the axis of the fitting model we assume that the vertical component of the magnetic field is non-zero, and that the azimuthal component of the field is zero (by definition).
We assume that $B_\theta$ and $B_z$  can both be expanded as a Taylor series in the radial direction about the flux-tube axis.
Under these assumptions, \citeauthor{1989GApFD..48..217F} showed that the Taylor-series expansion for $B_z$ can only contain even powers of $r$, whereas the expansion for $B_\theta$ can contain only odd powers of $r$.
It can then be shown that the Taylor-series expansion of the winding rate $\qmod$ in the radial direction about the flux-tube axis can contain only even powers of $r$, and $\qmod$ must be finite at the flux-tube axis.
We note that it is possible to construct a combination of $B_\theta$ and $B_z$, satisfying the above assumptions, such that $\qmod$ has a singularity in the flux-tube interior (\myeg{} at locations where $B_z=0$).
Thus, we restrict our attention to cases with $B_\theta$ and $B_z$ such that the winding rate is finite throughout the flux-tube interior (see Figure~\ref{egqw}(a) for a selection of example winding-rate profiles).
In this class of flux tubes, the current density has the following properties: its radial component is zero, the Taylor-series expansion for its azimuthal (vertical) component contains only odd (even) powers of $r$, and its azimuthal component vanishes at the flux-tube axis.
The Lorentz force only has a radial component, its Taylor-series expansion contains only odd powers of $r$, and it vanishes at the flux-tube axis.

Due to the assumptions outlined above, these fitting models may be too simple to provide a realistic representation of solar magnetic-flux tubes in general.
Nevertheless, it is reasonable to assume that these models could be a useful starting point for developing methods to estimate the winding rate of the magnetic field in solar magnetic-flux tubes; indeed, a similar model has been used by \inlinecite{Nandyetal2008} to estimate twist in solar active regions.
We use these fitting models because they are simple, generally current-carrying, generally have a non-zero Lorentz force (which is expected near the solar photosphere, \myeg{} \opencite{1995ApJ...439..474M}), and the winding rate can vary in the radial direction.
Furthermore, methods based on these simple flux-tube models can be tested and the influence of the assumptions built into the fitting model can be easily demonstrated.
One such assumption that we examine is the radial variation of the winding rate used in the fitting model.
These types of magnetic flux-tube models have been used in a couple of different solar contexts to understand the influence of different functional forms for the radial dependence of the winding rate.
For example, they have been used in theoretical studies to determine how the radial variation of the winding rate affects the susceptibility of these types of flux tubes to the kink instability (\myeg{} \opencite{1972SoPh...22..425R}; \opencite{HoodPriest1979}; \opencite{1983SoPh...88..163E}; \opencite{1990Ap+SS.166..289C}; \opencite{1990ApJ...361..690M}; \opencite{Vellietal1990}; \opencite{1998ApJ...494..840L}; \opencite{vanderLindenHood1998}, \citeyear{vanderLindenHood1999}; \opencite{1998A+A...333..313B}; \opencite{1997SoPh..172..249B}, \citeyear{Baty2001}).
They have also been used in numerical simulations of rising flux tubes to understand how the radial variation of the winding rate influences the emergence process (\myeg{} \opencite{1998ApJ...492..804E}; \opencite{1998MNRAS.298..433H}; \opencite{2008A+A...479..567M}).

\section{Estimating the Winding Rate in a Magnetic-Flux Tube}
\label{sec_fit}

To estimate the winding rate in a magnetic-flux tube we follow \inlinecite{Nandyetal2008} by formulating a classical fitting problem that seeks to minimize the discrepancy between the observed ratio
$\wrat = B_\theta / B_z$
and that produced by the fitting model,

\begin{equation}
\chi^2 = \frac{1}{N}   \sum_{i=1}^{N} \left( \wmod (r_i) - \wrat_{{\rm obs},i} \right)^2 \, ,
\label{chisq}
\end{equation}

\noindent
where 
$r_i$ is the radial coordinate of observation point $i$,
$N$ is the number of measurements used for the fitting,
$\wrat_{{\rm obs},i}$ is the ratio $B_\theta / B_z$ derived from the vector magnetic-field measurements at observation point $i$,
and $\wmod (r_i) = r_i \qmod (r_i)$ is the value of the ratio predicted by the fitting model at $r_i$.
For the fitting model, the Taylor-series expansion of $\wmod$ in the radial direction about the flux-tube axis can contain only odd powers of $r$, and $\wmod$ must vanish at the flux-tube axis (for some examples of $\wmod (r)$ see Figure~\ref{egqw}(b)).

The definition of $\chi^2$ given by Equation~(\ref{chisq}) could be modified in standard ways to include measurement uncertainties if they were available.
We use this definition because we are assuming (for demonstration purposes) that the measurements are free of error.
In practice the ratio $\wrat = B_\theta / B_z$ may not be the best quantity for fitting because $B_z$ is in the denominator and this may amplify noise and/or uncertainties in the measurements.
As mentioned previously, to avoid situations where $\wrat$ is undefined, we assume that $B_z$ is non-zero at all of the observation points for both the fitting model and the measurements.
We use the ratio $\wrat = B_\theta / B_z$ for the following reasons: 
it is sufficient for demonstrating our main findings (with error-free measurements),
it was employed by \inlinecite{Nandyetal2008}, and
it does not involve directly fitting for both components of the magnetic field, $B_\theta$ and $B_z$ (\myie{} a model for each of the components of the magnetic field is not required).

In what follows, we will determine the sensitivity of the winding rates estimated by this method to the assumptions built into the fitting model, using synthetic vector magnetic-field measurements generated with models for twisted magnetic-flux tubes.
In general, the objective of fitting a model to data in this way is to extract an underlying trend from the data and possibly to estimate some parameters related to the best-fit model.
To test the efficacy of the fitting method we focus on two quantities: the winding rate at the flux-tube axis and the radial average of the winding rate, 

\begin{equation}
\wind_{\rm av} = \frac{1}{b-a} \int_a^b \wind (r) dr \, ,
\end{equation}

\noindent
where $[ a, b ]$ is the range over which the average is taken.
We use the on-axis and average winding rates as metrics because they can be used to characterise the twist present in a magnetic-flux tube and both of these quantities may be useful for determining whether a flux tube is unstable to the kink instability (\myeg{} \opencite{1990ApJ...361..690M}; \opencite{Vellietal1990}; \opencite{1992SoPh..137..273R}; \opencite{Baty2001}).
We note that if sufficient  measurements of $\wrat_{{\rm obs},i}$ are available then these quantities may be approximated directly from the data (depending on the distribution of the observation points $r_i$) and, therefore, the fitting procedure may not be strictly necessary for error-free data.
However, we emphasise that the goal here is to examine how these quantities, as estimated from the best-fit model, are affected by the assumptions built into the fitting model.

\section{Fitting Models with a Fixed Radial Dependence for the Winding Rate}
\label{sec_sing}

To demonstrate how assumptions about the radial variation of the winding rate for the fitting model can affect estimates for the on-axis and average winding rates, we consider a class of fitting models with $\qmod(r) = \qmodamp f(r)$, 
where $\qmodamp$ is a parameter with dimensions of a winding rate (\myie{} radians per unit length) 
and $f(r)$ is a dimensionless function of $r$ that describes the radial variation of the winding rate (and satisfies the assumptions described in Section~\ref{sec_mod}, \myie{} the Taylor-series expansion of $f$ about $r=0$ contains only even powers of $r$).
In this section, we assume that the functional form of $f$ is fixed during the fitting procedure and that $f$ is independent of $\qmodamp$ (\myie{} $\partial f / \partial \qmodamp =0$).
Consequently, the minimization problem described by Equation~(\ref{chisq}) has only one free parameter [$\qmodamp$] that controls the amplitude of the winding rate for the fitting model.

By fixing the functional form of $f$ we are making an assumption about the radial variation of the winding rate.
This may be reasonable if observational or theoretical information is available to support this assumption, and it may be a useful starting point for developing fitting methods to estimate twist (\myeg{} \opencite{Nandyetal2008}).
To understand the implications of this assumption we construct error-free synthetic data using flux-tube models that satisfy the same assumptions as the fitting model
(\myie{} the magnetic field and winding rate do not vary in time or in the azimuthal or vertical directions, see Section~\ref{sec_mod} for details).
To this end, in the minimization problem described by Equation~(\ref{chisq}) we set $\wrat_{{\rm obs},i} = \qobsamp r_i g (r_i)$,
where $\qobsamp$ is a parameter with dimensions of a winding rate 
and $g(r)$ is a dimensionless function of $r$ that satisfies the same assumptions as $f (r)$.
Thus, for this exercise the only differences between the model used to construct the synthetic data and the fitting model are the values of $\qmodamp$ and $\qobsamp$, and the radial dependence of the winding rates described by $f$ and $g$.
To generate synthetic data we sample $\wrat_{{\rm obs},i}$ along a line at fixed $\theta$, which is appropriate because none of the quantities of interest vary in the azimuthal direction.

For this combination of synthetic data and fitting model, the minimization problem can be solved exactly.
It can be shown that the best-fit value for $\qmodamp$, corresponding to the minimum of $\chi^2$ defined by Equation~(\ref{chisq}), is $\qmodamp^\ast = c_\wrat \qobsamp$, where the asterisk  denotes the best-fit value and

\begin{equation}
c_\wrat = \frac{\sum_{i=1}^{N} r_i^2 f (r_i) g (r_i) }{\sum_{i=1}^{N} r_i^2 \left( f (r_i) \right)^2} \, .
\label{demosol2}
\end{equation}

\noindent
Consequently, the winding rate for the best-fit model is $\qmod^\ast(r) = c_\wrat \qobsamp f(r)$.
For synthetic data with $g (0) \neq 0$, the inferred value for the winding rate at the flux-tube axis is directly proportional to the true value, \myie{} $\wind_0^\ast =  \wind_{0,{\rm obs}} c_\wrat f(0)/g(0)$, where the $0$ in the subscript denotes the value at the axis.
The inferred and true average winding rates are also directly proportional, \myie{}

\begin{equation}
\wind_{\rm av}^\ast = \wind_{\rm av,obs} c_\wrat \frac{\int_a^b  f (r) dr}{\int_a^b g (r) dr} \, ,
\label{demosolav}
\end{equation}

\noindent
for synthetic data with $\int_a^b g (r) dr \neq 0$.
In each case, the constant of proportionality relating the inferred and true winding rates is determined by the mismatch between the radial variation of the winding rate assumed for the fitting model $f$ and that used to construct the synthetic data $g$, at the set of observation points $r_i$ and at the axis $r=0$ or over the averaging interval $[ a, b ]$.
However, it should be noted that, for cases with $f \neq g$ at some of the relevant locations, the constant of proportionality can theoretically be unity (meaning that the inferred winding rate matches the true value).

To demonstrate the implications of these results, we test a fitting model with a uniform winding rate, \myie{} $f(r)=1$.
A similar fitting model was used by \inlinecite{Nandyetal2008}. 
The difference between this approach and that taken by \citeauthor{Nandyetal2008} is that their fitting model has two free parameters, and they did not require $w=0$ at $r=0$.
We construct three different synthetic data sets using  different choices for the radial variation of the winding rate (see, \myeg, Figure~\ref{egqw}):
$g(r)=(1-(r/R)^2)^2$, 
$g(r) = 1 + 2 (r/R)^2 - 3 (r/R)^4$, and
$g(r) = \exp ( -(r/R)^2)$, 
where $R$ is the radius of the flux tube. 
These synthetic winding-rate profiles are chosen for the following reasons:
they satisfy the requirement that the Taylor-series expansion of winding rate about $r=0$ contains only even powers of $r$, 
they are single signed in the interval $0 \le r \le R$, 
and profiles like these have been used previously by \inlinecite{HoodPriest1979}, \inlinecite{1990ApJ...361..690M}, \inlinecite{1998ApJ...494..840L}, \inlinecite{1998A+A...333..313B}, and \inlinecite{Baty2001}.
Synthetic measurements are generated at two hundred equally spaced observation points in the interval $0 \le r_i \le  R$, all of which are used for the fitting, and the average winding rate is calculated for the interval $[ 0,  R ]$.

\begin{table}
\caption{Results for tests of a fitting model with a uniform winding rate [$\qmod (r) = \qmodamp$] applied to several different synthetic data sets with $\wrat_{{\rm obs},i} = \qobsamp r_i g (r_i)$.
}
\label{demo2tab}
\begin{tabular}{lccc}
\hline
Synthetic data generated with $g(r)$: & $ \wind_0^\ast /  \wind_{0,{\rm obs}} $ &  $ \wind_{\rm av}^\ast  / \wind_{\rm av,obs} $ & $\chi^2 / (R^2 \qobsamp^2)$ \\[1mm]
\hline
$g (r) = (1-(r/R)^2)^2$                         & 0.227 & 0.425  & 0.0196 \\[1mm]
$g (r) = 1 + 2 (r/R)^2 - 3 (r/R)^4 $            & 0.907 & 0.851  & 0.0557 \\[1mm]
$g (r) = \exp ( -(r/R)^2)$                      & 0.567 & 0.759  & 0.00806 \\[1mm]
\hline
\end{tabular}
\end{table}

The results for these tests (see Table~\ref{demo2tab}) show that the fitting model with a uniform winding rate does not retrieve the true value for the winding rate at the flux-tube axis or the average winding rate for any of the synthetic data sets; that is, the ratios $ \wind_0^\ast /  \wind_{0,{\rm obs}} $ and $ \wind_{\rm av}^\ast  / \wind_{\rm av,obs} $ both deviate substantially from unity (in some cases by several tens of percent).
In other experiments with different combinations of $f$ and $g$ we find similar outcomes (results not shown).
For the best-fit model, $\chi^2$ is directly proportional to $\qobsamp^2$, \myie{}

\begin{equation}
\chi^{2 \, \ast} = \frac{\qobsamp^2}{N}  \sum_{i=1}^{N} r_i^2  \left( c_\wrat  f (r_i) -  g (r_i) \right)^2 \, ,
\label{chisqsing}
\end{equation}

\noindent
and the results in Table~\ref{demo2tab} show that smaller values of $\chi^2$ do not necessarily correspond to more accurate winding-rate estimates.

For the tests in Table~\ref{demo2tab} the fitting model underestimates the magnitudes of both the on-axis and average winding rates.
In other experiments we find cases where the fitting model with $f(r)=1$ overestimates the magnitude of the winding rates.
Whether a particular fitting model $f$ underestimates or overestimates the magnitude of the winding rate depends on the functional forms of $f$ and $g$, and on the locations of the observation points $r_i$.

The fitting model tested in Table~\ref{demo2tab} retrieves the correct sign for both the on-axis and average winding rates.
This is expected for this combination of synthetic data and fitting model, and can be understood by inspecting the solutions for the best-fit winding rates presented above.
At the axis, the inferred and true winding rates will have the same sign if both $f$ and $g$ satisfy the following conditions: they are single signed for all of the observation points $r_i$  and have the same sign at the axis that they have at the observation points $r_i$.
Likewise, the inferred and true average winding rates will have the same sign if both $f$ and $g$ satisfy the following conditions: they are single signed for all of the observation points $r_i$ and in the averaging interval $[a,b]$, and they have the same sign in the interval $[a,b]$ that they have at the observation points $r_i$.

\section{Fitting Methods that Infer the Radial Dependence of the Winding Rate}
\label{sec_multi}

In the previous section, we demonstrated how an error can be introduced into the estimates for the winding rates when the radial variation of the winding rate assumed for the fitting model is fixed and does not match that in the flux-tube model used to generate the synthetic measurements.
In this section we show how this problem can be addressed by inferring the radial variation of the winding rate during the fitting procedure (see, \myeg, Chapter~15.4 of \opencite{1992nrfa.book.....P}).
To this end, we assume that the winding rate in the fitting model can be approximated as

\begin{equation}
\qmod (r) = \sum_{j=1}^{n_b} \qbamp_j \basis_j (r) \, ,
\label{basis}
\end{equation}

\noindent
where $\basis_j (r)$ are dimensionless basis functions,
$\qbamp_j$ are amplitudes with dimensions of a winding rate,
and $n_b$ is the number of basis functions used in the approximation.
The basis functions [$\basis_j$] must be chosen such that the radial variation of the winding rate satisfies the conditions discussed in Section~\ref{sec_mod} (\myie{} 
the Taylor-series expansion for the winding rate [$\qmod$] in the radial direction about the flux-tube axis contains only even powers of $r$, and $\qmod$ is finite throughout the flux-tube interior).

For this approach, the aim is to find the values for the set of amplitudes [$\qbamp_k$] for a given set of basis functions [$\basis_k$] that correspond to the minimum of $\chi^2$ (as defined by  Equation~(\ref{chisq})).
The resulting system of linear equations (for $\partial \chi^2 / \partial \qbamp_k =0$, $k=1, \ldots, n_b $) is:

\begin{equation}
\sum_{j=1}^{n_b} A_{kj} \qbamp_j = C_k \, , 
\label{linsys}
\end{equation}

\noindent
where

\begin{equation}
A_{kj}= \sum_{i=1}^{N} r_i^2 \basis_k (r_i) \basis_j (r_i) \, , \quad \mbox{and} \quad C_k= \sum_{i=1}^{N} r_i \basis_k (r_i) \wrat_{{\rm obs},i} \, ,
\label{abkj}
\end{equation}

\noindent
which can be solved using standard methods of linear algebra (provided that the matrix $A_{kj}$ is not singular or ill-conditioned, see, \myeg, Chapter~2 of \opencite{1992nrfa.book.....P}); in the examples that follow we use singular value decomposition and monitor the condition number of the matrix $A_{kj}$ to ensure that it is not ill-conditioned.

To test this approach we follow the same procedure that we used in Section~\ref{sec_sing}.
That is, we construct error-free synthetic data using a flux-tube model that satisfies the same assumptions as the fitting model (\myie{} the magnetic field and winding rate do not vary in time or in the azimuthal or vertical directions) by setting $\wrat_{{\rm obs},i} = \qobsamp r_i g (r_i)$.
For this type of synthetic data it is evident from Equations~(\ref{linsys})\,--\,(\ref{abkj}) that the best-fit amplitudes $\qbamp_j$ are directly proportional to $\qobsamp$.
Therefore, the best-fit on-axis and average winding rates are directly proportional to the respective true values (for synthetic data with $g (0) \neq 0$ and $\int_a^b g (r) dr \neq 0$), and  $\chi^2$  for the best-fit model is directly proportional to $\qobsamp^2$.
For demonstrating this approach we use even-order polynomials for the basis functions [$\basis_j (r) = (r/R)^{2 (j-1)}$] which are a convenient choice  because they satisfy the required properties for the Taylor-series expansion of the winding rate about the flux-tube axis.
We use two different values for $n_b$ [$n_b=2$ and $n_b=3$].
We note that polynomials may not be the best choice for the basis functions in practice, but determining the optimal type of basis functions is beyond the scope of this investigation.
Synthetic measurements are generated at two hundred equally-spaced observation points in the interval $0 \le r_i \le  R$ using $g(r) = \exp ( -(r/R)^2)$.
We use this synthetic winding-rate profile because its Taylor-series expansion is infinite (\myie{} it cannot be represented exactly by a low-order polynomial approximation).
The average winding rate is calculated for the interval $[ 0 , R ]$.

\begin{table}
\caption{Results for tests of several fitting models applied to synthetic data constructed with $\wrat_{{\rm obs},i} = \qobsamp r_i \exp ( -(r_i/R)^2)$}
\label{toytab}
\begin{tabular}{lccc}
\hline
Fitting model   & $ \wind_0^\ast /  \wind_{0,{\rm obs}} $ &  $ \wind_{\rm av}^\ast  / \wind_{\rm av,obs} $ & $\chi^2 / (R^2 \qobsamp^2)$\\[1mm]
\hline
$\qmod (r) = \qmodamp$                                        & 0.567 & 0.759  & 0.00806 \\[1mm]
$\qmod (r) = \qmodamp (1-(r/R)^2)^2$                          & 1.507 & 1.076  & 0.0320 \\[1mm]
$\qmod (r) = \qmodamp  ( 1 + 2 (r/R)^2 - 3 (r/R)^4 )$         & 0.571 & 0.815  & 0.00764 \\[1mm]
$\qmod (r) = \qbamp_1  + \qbamp_2 (r/R)^2$                    & 0.920 & 0.970  & 0.000132 \\[1mm]
$\qmod (r) = \qbamp_1  + \qbamp_2 (r/R)^2 + \qbamp_3 (r/R)^4$  & 0.991 & 0.998  & 9.46$\times 10^{-7}$ \\[1mm]
\hline
\end{tabular}
\end{table}

We compare the results of this approach with those from several fitting models that assume a fixed radial dependence for the winding rate (as described in Section~\ref{sec_sing}).
For this synthetic data set, the results for these tests (see Table~\ref{toytab}) show that  both the on-axis and average winding rates retrieved by the best-fit polynomial basis-function approximations are more accurate than those retrieved with the fitting models with a fixed radial variation for the winding rate.
The corresponding $\chi^2$ values are also much smaller for the best-fit polynomial basis-function approximations.
For low-order polynomial basis-function approximations, we find that the winding-rate estimates and corresponding $\chi^2$ values improve as $n_b$ increases.
In other experiments with the same set of basis functions, but different types of synthetic data, we find a similar pattern (see Appendix~\ref{sec_alpha}).
For fitting methods with a fixed assumption for the radial variation of the winding rate (as described in Section~\ref{sec_sing}), the results in Table~\ref{toytab} show that different assumptions for the radial variation of the winding rate can produce different estimates for the winding rates (given the same synthetic data).

\section{Testing Other Fitting-Model Assumptions}
\label{sec_assump}

In the previous section, we demonstrated that fitting methods that infer the radial variation of the winding rate can retrieve more accurate estimates for the on-axis and radially-averaged winding rates than methods that make a fixed assumption about the radial variation of the winding rate (if the assumed radial variation for the fitting model does not match that used to construct the synthetic data).
Those tests used synthetic data constructed with models that satisfy the same assumptions as the fitting model, except for the assumption regarding the radial dependence of the winding rate, which was considered an unknown.
As discussed above, the fitting models used here are very simple and may not  provide a realistic representation for twisted solar magnetic-flux tubes in general.
The purpose of this section is to determine whether the fitting methods developed up to this point, which are based on simplified models for twisted magnetic-flux tubes, can be expected to retrieve accurate winding-rate estimates for synthetic data constructed with models that {\it do not} satisfy all of the assumptions built into the fitting model.
To re-iterate, the assumptions built into the fitting model include:
the flux tube is static and isolated,
the flux-tube axis is straight and parallel to the $z$-direction,
the location of the flux-tube axis is known,
the fitting model is axisymmetric (\myie{} the magnetic field and winding rate in the fitting model do not vary in the $\theta$-direction),
and the magnetic field and the winding rate do not vary in the $z$-direction.
In the Appendix we conduct a series of experiments that document the effects of  some of these assumptions on the winding-rate estimates. For the sake of brevity, here we briefly summarise the main findings from each experiment, before applying the fitting method to synthetic data generated with a twisted toroidal magnetic-flux-tube model.

Regarding variations of the magnetic field in the vertical direction for the model used to construct the synthetic data (see Appendix~\ref{sec_alpha} for details and examples), if the observed flux tube is axisymmetric, with an axis that is straight and vertical, we find that reasonable winding-rate estimates can be retrieved for the plane where the magnetic field is measured by the fitting methods that infer the radial variation of the winding rate.
The winding rate at the axis of an axisymmetric magnetic-flux tube, with an axis that is straight and vertical, does not vary along the length of the axis (\myeg{} \opencite{1989GApFD..48..217F}).
Therefore, an estimate for the winding rate at the flux-tube axis retrieved using measurements at one height can be applied along the entire axis of the flux tube, provided that the observed flux tube is indeed axisymmetric, with an axis that is straight and vertical.
On the other hand, away from the flux-tube axis, if the magnetic field varies with height we find that the winding rate determined for a given radial location (or field line) at one height may not correspond to the winding rate at the same radial location (or field line) at a different height.
Thus, the average winding rate found by fitting methods using measurements at one height cannot generally be applied to other heights, without additional information.

Regarding variations of the magnetic field in the azimuthal direction for the model used to construct synthetic data (see Appendix~\ref{sec_ddthne0}), it is clear that fitting methods that use axisymmetric fitting models cannot retrieve the azimuthal dependence of the winding rate.
With that in mind, we find that fitting methods that use axisymmetric fitting models can retrieve reasonable estimates for the azimuthally-averaged, radially-averaged winding rate and the on-axis winding rate for synthetic measurements constructed with a  magnetic field that varies in both the azimuthal and radial directions (but not in the vertical direction).
For example,  we find that the relative discrepancy between the inferred and true winding rates is of order 10\,\% when the winding rate varies in the azimuthal direction by up to a factor of three in some parts of the observed flux tube.

In Appendix~\ref{sec_axisloc} we find that an error in the location of the flux-tube axis generally produces an error in the winding rates retrieved by fitting methods, in the absence of other sources of error.
The relative error in the winding-rate estimates is of order 10\,\% when the error in the location of the flux-tube axis is of order 10\,\% of the flux-tube diameter, for the cases we have examined.
We also find that the error in the winding-rate estimates increases as the error in the location of the flux-tube axis increases.

\subsection{Testing the Fitting Methods with a Twisted Toroidal Magnetic-Flux Loop}
\label{sec_tor}

To test the efficacy of the fitting methods applied to solar-like magnetic fields we employ the twisted, toroidal, magnetic-flux-loop model used to drive the simulations of \citeauthor{fan03}~(\citeyear{fan03}, \citeyear{fan04}) and \inlinecite{2010ApJ...720..233C}.
This model is useful for constructing synthetic data because it qualitatively resembles some twisted solar magnetic structures (depending on the choice of model parameters, see, \myeg, \opencite{kink}),
the winding rate of the field lines about the flux-loop axis is well defined, 
and the model violates several of the assumptions built into the fitting model (this partly depends on the choice of model parameters).
Moreover, the simulation results of \citeauthor{fan03}~(\citeyear{fan03}, \citeyear{fan04}) have been used for  ``blind'' testing methods for analysing vector-field measurements, such as for twist estimation (\myeg{} \opencite{kink}) and azimuthal ambiguity resolution (\myeg{} \opencite{2006SoPh..237..267M}; \opencite{2007ApJ...654..675L}; \opencite{2009SoPh..260..271C}).

To describe this magnetic field we use a spherical coordinate system $(\rw, \pw, \tw)$, where 
$\tw$ is the angle measured counterclockwise from the positive $\xw$-axis to the position vector in the  $\yw$--$\xw$-plane,
$\pw$ is the polar angle measured from the positive $\zw$-axis to the position vector, and 
$\rw$ is the distance from the position $(x, y, z)$ to the origin (\myie{} $\rw^2 = x^2 + y^2 + z^2$).
The flux-loop axis is circular, lies in the $\yw$--$\xw$-plane  (see Figure~1 of \citeauthor{fan03} \citeyear{fan03}, \citeyear{fan04}), and is located at  $\pw = \pi/2$ and $\rw = \rl$.
The magnetic field is given by

\begin{equation}
\B = \curl \left[ \frac{A \left( \rw, \pw \right)}{ \rw \sin \pw } \mathbf{e_\tw} \right] + B_\tw \left( \rw, \pw \right)  \mathbf{e_\tw} \, ,
\label{fg}
\end{equation}

\noindent
where

\begin{equation}
A \left( \rw, \pw \right) = \frac{1}{2} \ql \al^2 \bt \exp \left[ - \frac{ \varpi^2 \left( \rw , \pw \right) }{\al^2} \right] \, , 
\quad
B_\tw \left( \rw , \pw \right) = \frac{ \al \bt }{ \rw \sin \pw } \exp \left[ - \frac{ \varpi^2 \left( \rw , \pw \right) }{\al^2} \right] \, , 
\label{fg2}
\end{equation}

\noindent
$\bt$ is a magnetic-field strength,
$\varpi^2 \left( \rw , \pw \right) = \rw^2 + \rl^2  - 2 \rw \rl \sin \pw$ is the squared distance from the position $(x, y, z)$ to the flux-loop axis at constant $\tw$,
$\al$ is a length scale,
and $\ql$ is a dimensionless parameter that is related to the rate at which the field lines wind around the flux-loop axis.
As in \citeauthor{fan04}, we truncate the magnetic field to zero for $\varpi \ge 3 \al$.

To understand some features of this magnetic field, it is helpful to consider a coordinate system $( \varpi, \psi, \tw )$
that results from two cylindrical coordinate transformations:

\begin{equation}
\rc^2 = \yw^2 + \xw^2  \, ,
\quad 
\rc \sin \tw = \yw \, ,
\quad 
\rc \cos \tw = \xw \, ,
\end{equation}

\noindent
and

\begin{equation}
\varpi^2 = \mysc^2 + \zw^2  \, ,
\quad  
\varpi \cos \psi =  \mysc  \, ,
\quad 
\varpi \sin \psi = \zw \, ,
\end{equation}

\noindent
where $\varpi$ and $\tw$ are as defined above,  $\mysc = \rl - \rc $, and $\psi$ is the angle about the flux-loop axis measured counterclockwise from the positive $\mysc$-axis to the position vector in the $\mysc$--$\zw$-plane at constant $\tw$.
In this coordinate system it can be shown that $B_\varpi=0$ and  $B_\psi= (\ql/\al) \varpi B_\tw$ (\myeg{} \opencite{2009A+A...503..999H}, \opencite{2009A+A...507..995M}).
Thus, the winding rate of the field lines about the flux-loop axis, per unit length along the axis, is $B_\psi / ( \varpi B_\tw ) = \ql/\al$, which is constant, yet $B_\psi$ and $B_\tw$ vary in both the $\varpi$- and $\psi$-directions  but not in the $\tw$-direction.

To construct synthetic data we use the same parameter values used by \citeauthor{fan03}~(\citeyear{fan03}, \citeyear{fan04}).
We sample the magnetic field at a set of discrete locations on an $x$\,--\,$y$-plane at constant height $z_0$, with a grid spacing of $0.00625 \, \fgl$ in both the $x$- and $y$-directions (where $\fgl$ is a length scale, corresponding to the size of the domain in the $\zw$-direction).
For the number of observation points in the $x$- and $y$-directions we set $n_x=240$ and $n_y=160$.
We set $\rl = 0.375 \, \fgl$ and $\al = 0.1 \, \fgl$.
At the flux-loop axis the only non-zero component of the magnetic field is parallel to the axis (\myie{} $B_\psi=0$  at $\varpi  =  0$), and it has a magnitude of $\al \bt / \rl$.
We choose the value of $\bt$ such that the magnetic-field strength at the axis is $2.4$~kG, and the positive-polarity footpoint is located in the $x < 0$ half-space.
A typical value for the winding rate is chosen by making the following considerations:
The twist $\twist$ for field lines in the vicinity of the flux-loop axis, over the length of the semi-circle coinciding with the axis, is given by $ \twist = \wind \myl = \ql \myl / \al$, where $\myl =  \pi \rl$.
\citeauthor{fan03} use $\ql=-1$ and, thus, the twist over the semi-circular axis in that case is $-1.875 \times 2 \pi$~radians.
The critical twist typically quoted for the onset of the kink instability is $| \wind \myl | \approx 2 \pi$~radians (\myeg{} \opencite{1972SoPh...22..425R}; \opencite{HoodPriest1979}; \opencite{1983SoPh...88..163E}; \opencite{1990ApJ...361..690M}; \opencite{Vellietal1990}; \opencite{1998ApJ...494..840L}; \opencite{vanderLindenHood1998}, \citeyear{vanderLindenHood1999}; \opencite{1998A+A...333..313B}; \opencite{Baty2001}; \opencite{fan03}, \citeyear{fan04}; \opencite{tor04}).
To test the performance of the fitting methods for a range  of relevant winding rates, we construct 20 synthetic vector magnetograms with 20 equally-spaced values for $\ql$ in the range $- 1 \le \ql \le  1$; the case where the winding rate is zero is not considered. 
For each synthetic magnetogram we analyse the positive-polarity footpoint of the flux loop as observed on the plane $z=z_0$.
For the fitting models we set  $R = 3 \al $ (\myie{} the radius of the flux tube).
At each observation point with $B_z > 100$~G and $r < R$  we compute the ratio $\wrat = B_\theta / B_z$, using the known location of the flux-loop axis on the $z=z_0$ plane to determine the radial coordinate $r$, the azimuthal angle $\theta$, and $B_\theta$.
We calculate the average winding rate over the interval used for the fitting (\myie{} $0 \le r \le \max (r_i)$).
We note that by using the known values for the radius of the flux loop and the location of the flux-loop axis we are intentionally giving the fitting methods the best possible chance for success, although in practice these parameters are not known.

\begin{figure}[ht]
\includegraphics[width=0.32\textwidth]{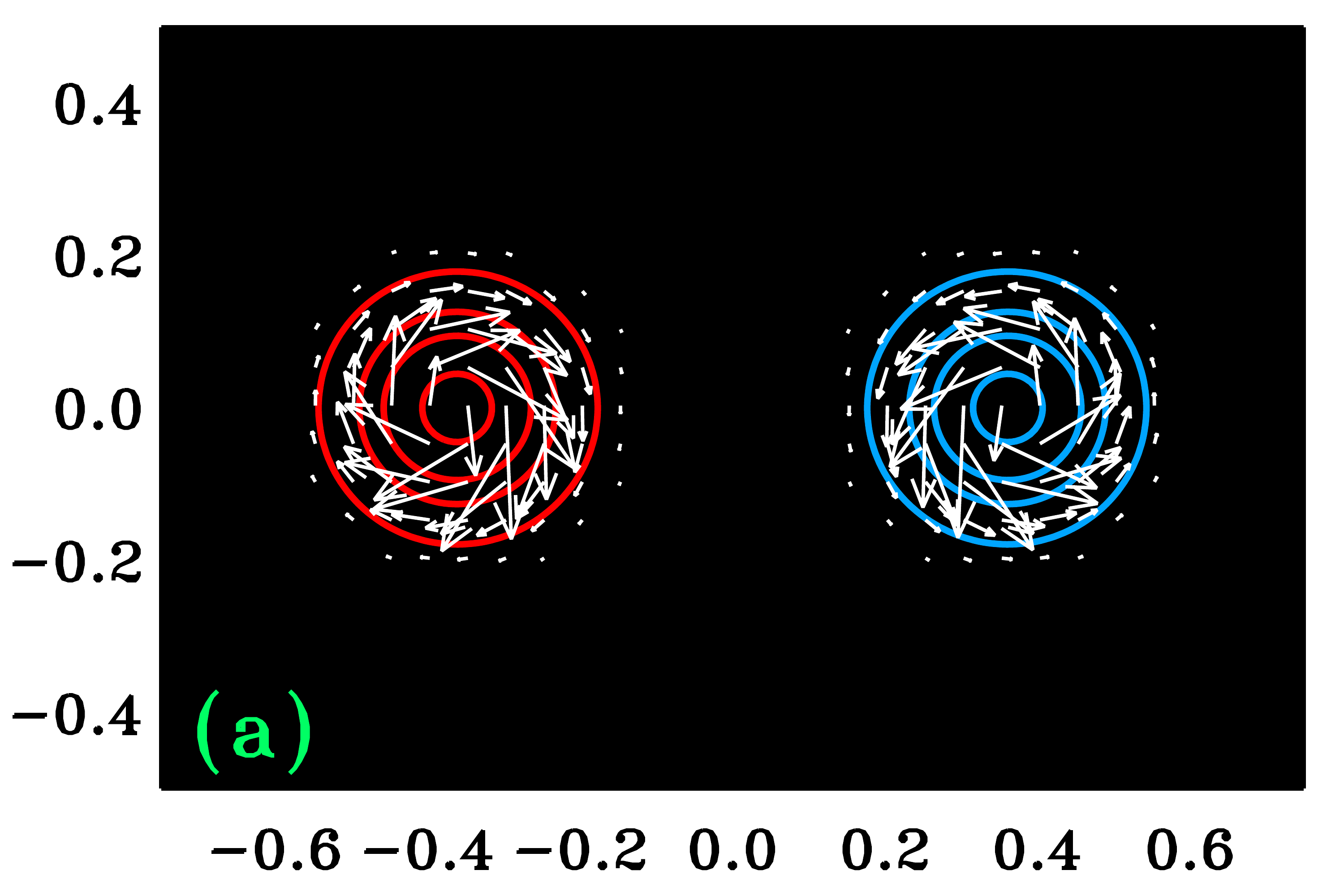}\hfil\includegraphics[width=0.32\textwidth]{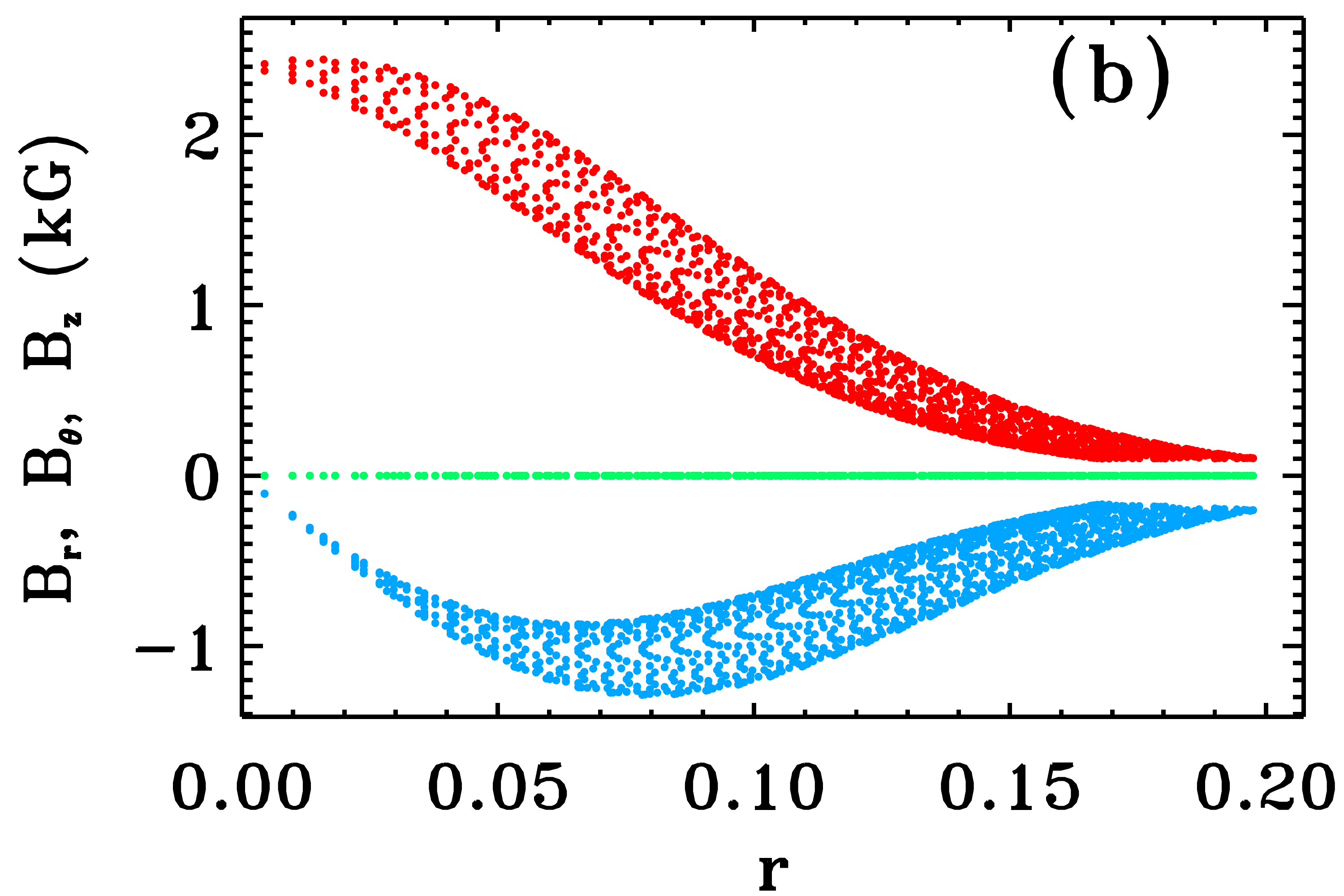}\hfil\includegraphics[width=0.32\textwidth]{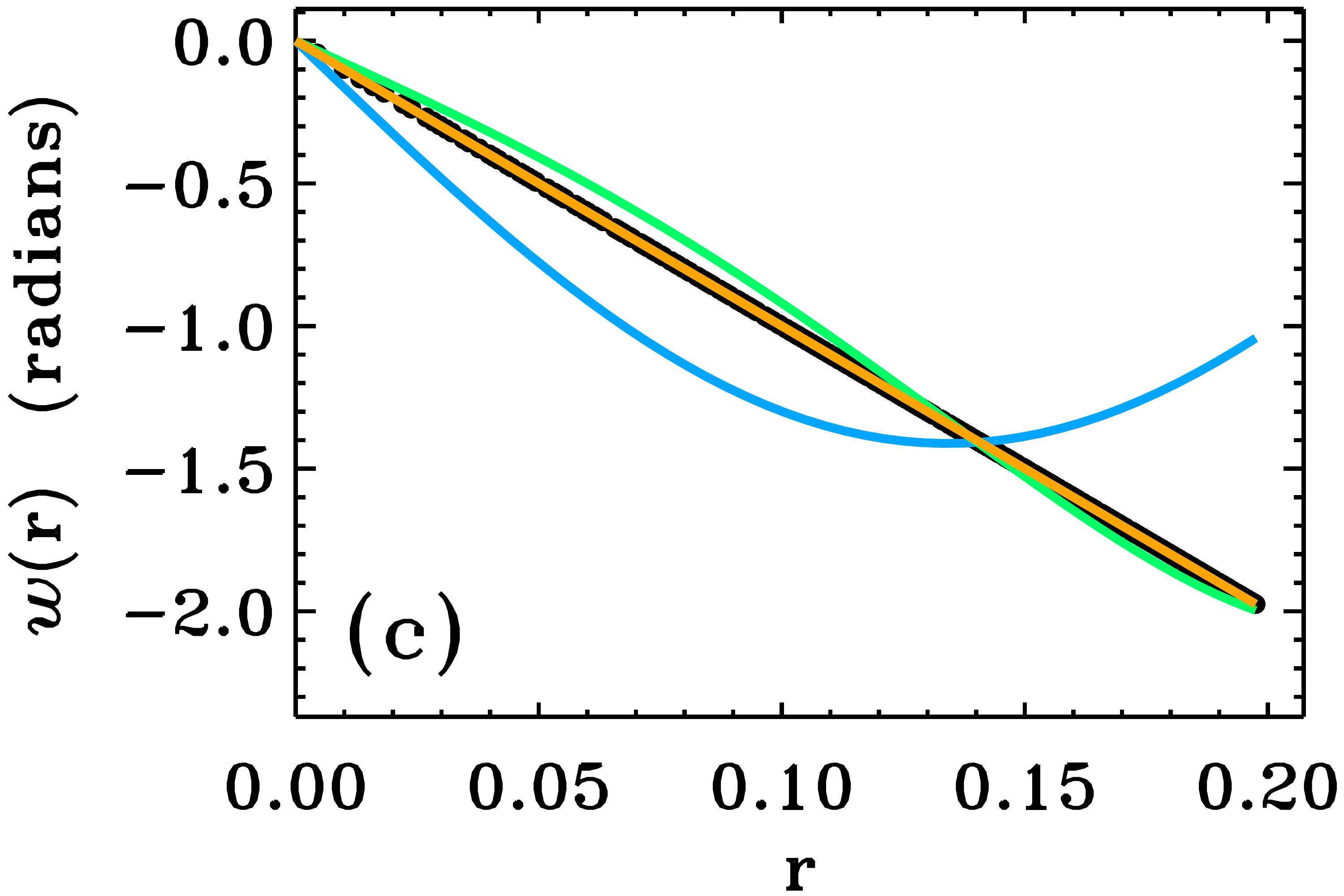}
\\[2mm]
\includegraphics[width=0.32\textwidth]{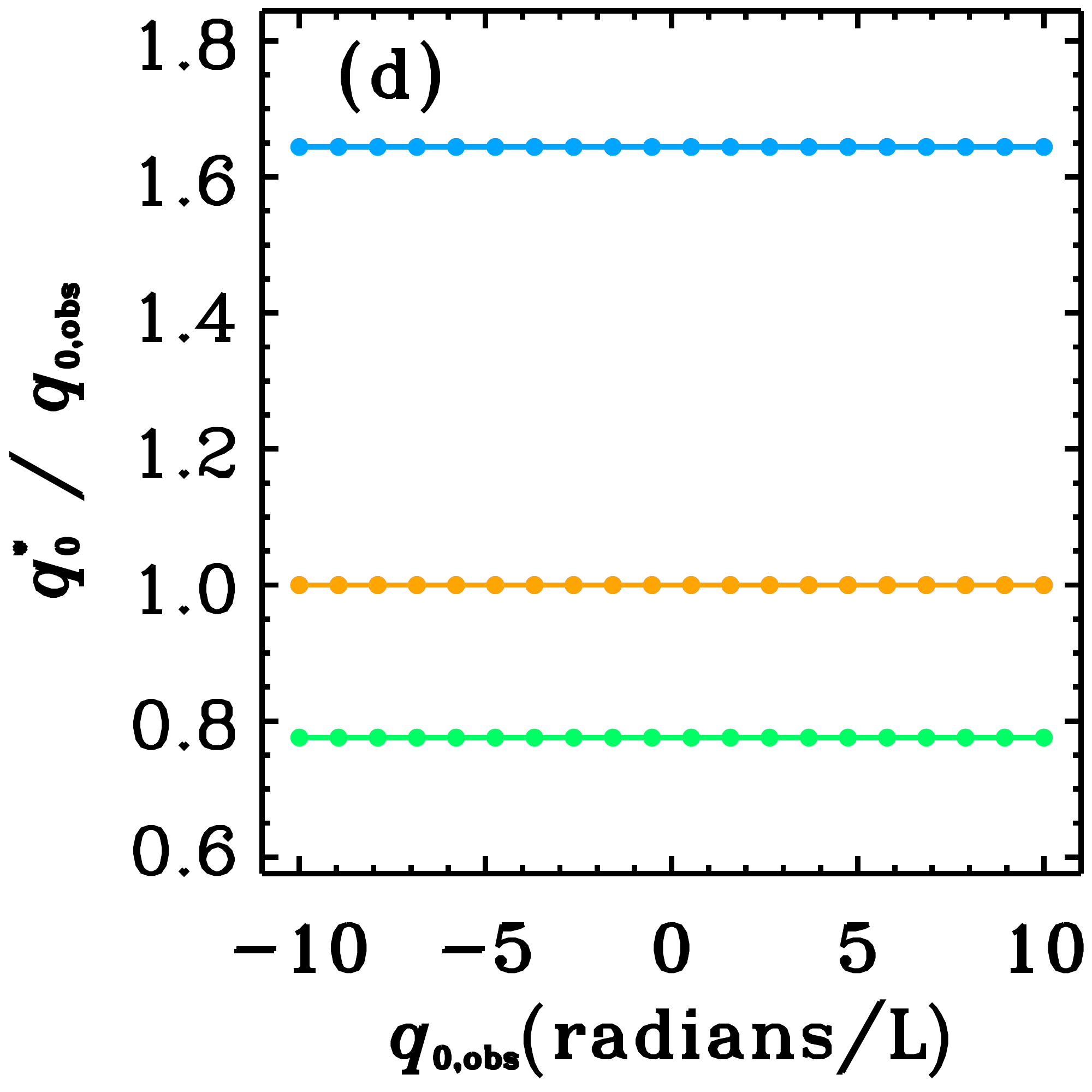}\hfil\includegraphics[width=0.32\textwidth]{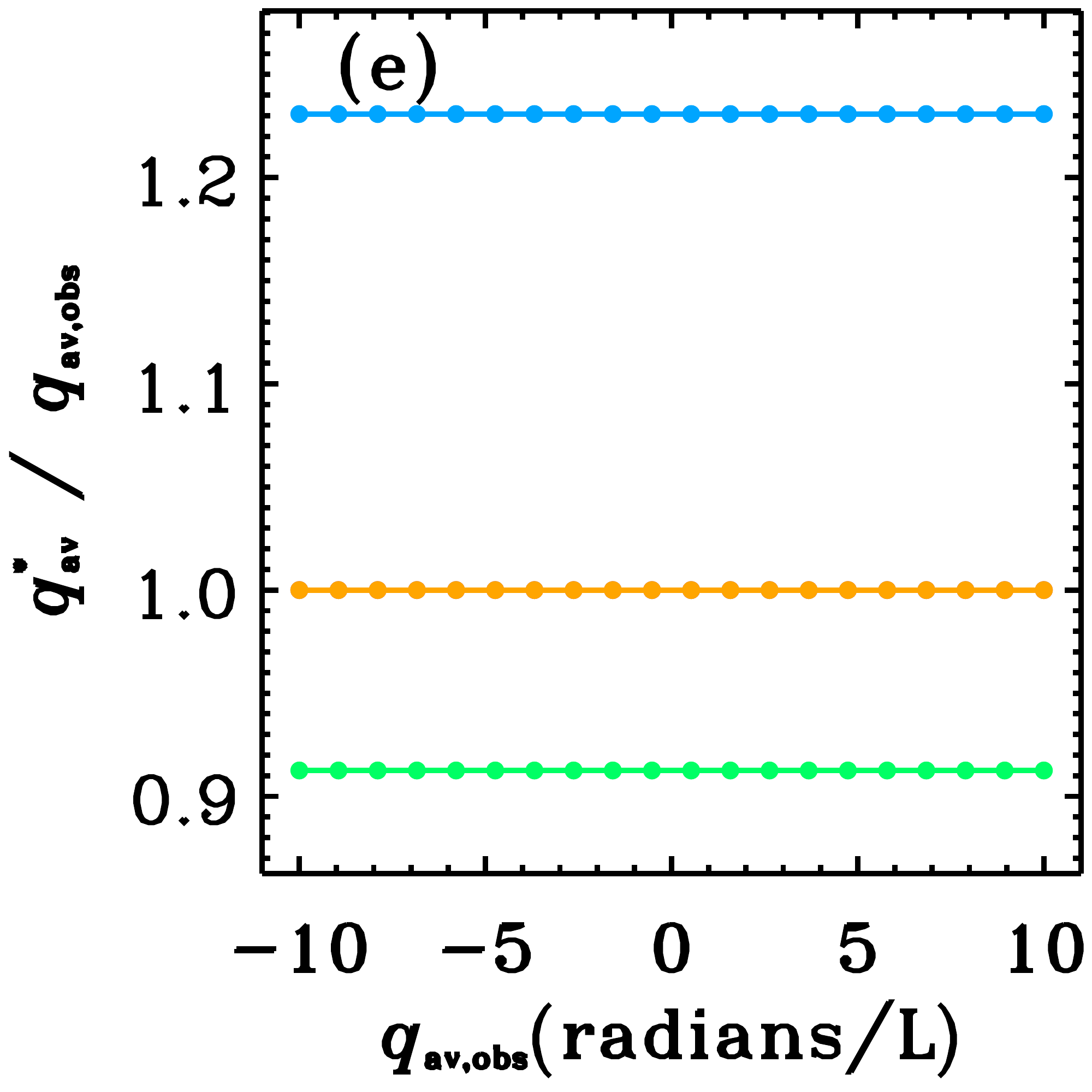}\hfil\includegraphics[width=0.32\textwidth]{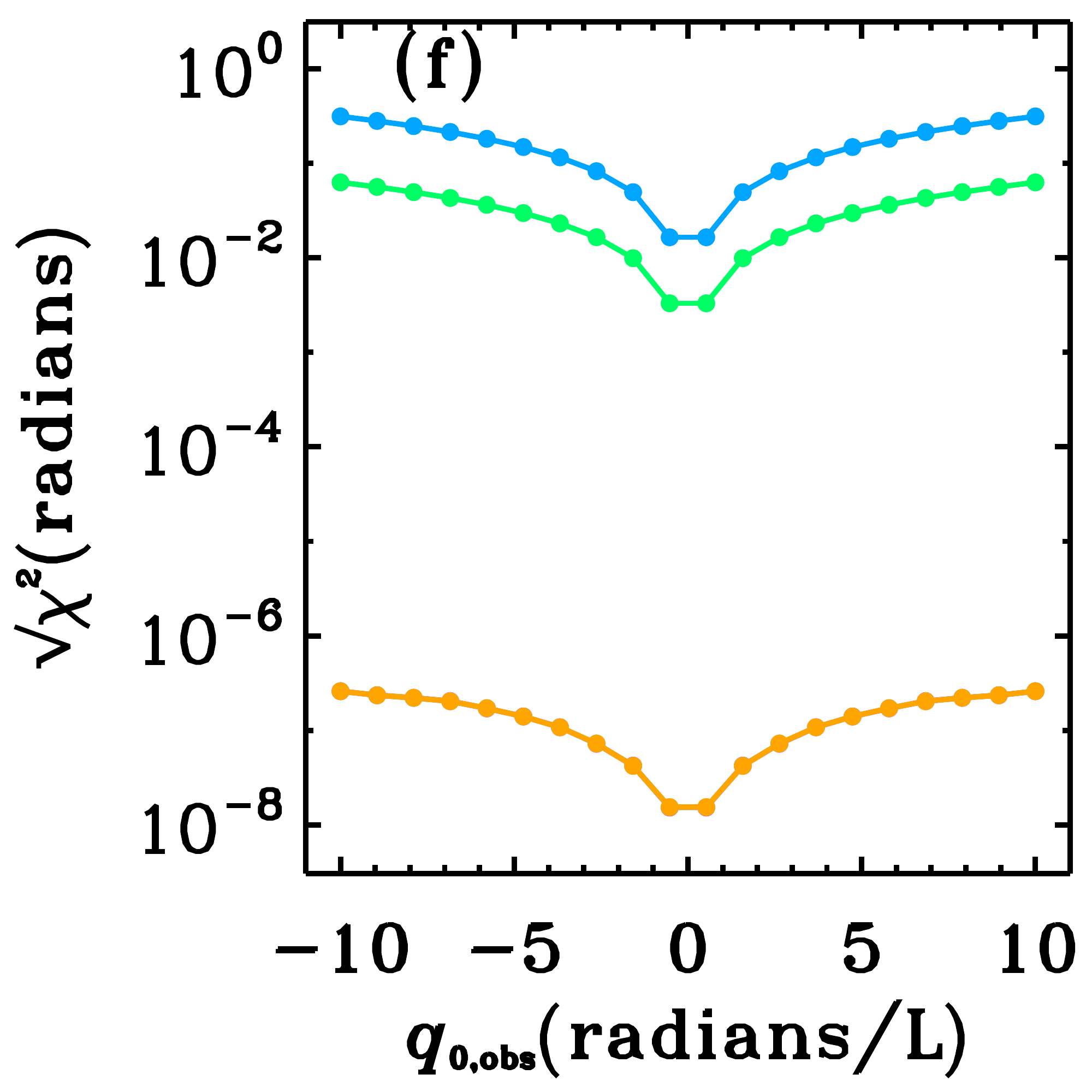}
\caption{
(a) Vector magnetic field at $z_0=0$ for the toroidal flux loop  with $(\ql/\al) =-10 $~radians $ \fgl^{-1}$.
Positive (negative) vertical magnetic field  [$B_z$] is indicated by red (blue) contours at 100, 500, 1000, 2000~G.
Horizontal magnetic field is plotted at every eighth pixel, with magnitude proportional to arrow length (the maximum horizontal field strength shown is approximately     1286~G); arrows with a horizontal magnetic-field strength less than 25~G are not shown.
(b) Inferred cylindrical components of the magnetic field [$B_r$, $B_\theta$, $B_z$] as a function of radial distance from the flux-loop axis [$r$], for the positive-polarity footpoint of the  magnetic field shown in (a).
The red, blue, and green points correspond to $B_z$, $B_\theta$, and $B_r$, respectively (note: $B_r=0$ for $z_0=0$).
(c) The ratio, $\wrat = B_\theta / B_z$, as a function of radial distance from the flux-loop axis [$r$].
The black points are the measurements derived from the synthetic data in (a) and (b).
The purple curve is the best-fit case for a fitting model with $\qmod (r) = \qmodamp$,
the blue curve is for $ \qmod (r) = \qmodamp (1-(r/R)^2)^2$,
the green curve is for $\qmod (r) = \qmodamp ( 1 + 2 (r/R)^2 - 3 (r/R)^4 )$,
the gray curve is for $\qmod (r) =  \qbamp_1  + \qbamp_2 (r/R)^2$,
and the orange curve is for $\qmod (r) =  \qbamp_1  + \qbamp_2 (r/R)^2 + \qbamp_3 (r/R)^4$;
note that the purple, gray, and orange curves are effectively indistinguishable here but can be discerned in subsequent figures.
(d)
The ratio of the inferred and true winding rates at the flux-tube axis [$\wind_0^\ast / \wind_{0,{\rm obs}}$] as a function of the true winding rate at the axis [$\wind_{0,{\rm obs}}$] for the fitting models in (c).
In these tests we use the toroidal flux-loop model of Fan and Gibson~(2003, 2004) to generate twenty synthetic data sets with a range of winding rates $\ql/\al$ (see text).
Each point represents the result from a single fitting experiment.
Note that the curves joining the points are only included as a guide and that the purple, gray, and orange curves are effectively indistinguishable here but can be discerned in subsequent figures.
The colors are the same as in (c).
(e) 
Same as (d) except for the corresponding ratio of the inferred and true average winding rates  [$\wind_{\rm av}^\ast / \wind_{\rm av,obs}$] as a function of the true average winding rate [$\wind_{\rm av,obs}$], with the average taken over the range used for the fitting.
(f) The corresponding value of $\sqrt{ \chi^2 }$ (see Equation~(\ref{chisq})) as a function of $\wind_{0,{\rm obs}}$.
}
\label{fgz0}
\end{figure}

Figure~\ref{fgz0} shows results when the magnetic field is sampled on the plane $z_0=0$.
This plane is special in the sense that the cylindrical directions [$r$, $\theta$, and $z$] are locally equivalent to the directions $\varpi$, $\psi$, and $\tw$, respectively (\myie{} the axis of the flux loop is parallel to the $z$-direction for $z_0=0$).
Thus, on this plane the radial component of the magnetic field [$B_r$] is zero.
On the other hand, $B_\theta$ and $B_z$ 
are generally non-zero and vary in both the $\theta$- and $r$-directions; that is, for a given value of $r$ the inferred values of $B_\theta$ and $B_z$ are not single-valued (evident as the scatter in Figure~\ref{fgz0}(b)).
Hence, the assumption that the magnetic field is axisymmetric is violated.
However, the inferred ratio $\wrat$ is single-valued for a given value of $r$ on the $z_0=0$ plane.
The fitting model that assumes a uniform winding rate profile can therefore retrieve the correct values for both the on-axis and average winding rates to a high degree of accuracy, despite some of the fitting-model assumptions being violated.
The same is also true for the best-fit polynomial basis-function approximations.
On the other hand, the fitting models that assume a fixed, non-uniform winding rate do not retrieve the correct winding rates.
These results indicate that the winding rates retrieved by the fitting methods are not  sensitive to variations of the field in the $\theta$-direction, provided that the winding rate itself does not vary in the $\theta$-direction; we acknowledge that this situation may be unlikely in practice.
These results also indicate that the winding-rate estimates are not strongly sensitive to the curvature of the flux-loop axis on the plane $z_0=0$;  in experiments with smaller values of $\rl$ we find results  similar to those in Figure~\ref{fgz0} (results not shown) .

\begin{figure}[ht]
\includegraphics[width=0.32\textwidth]{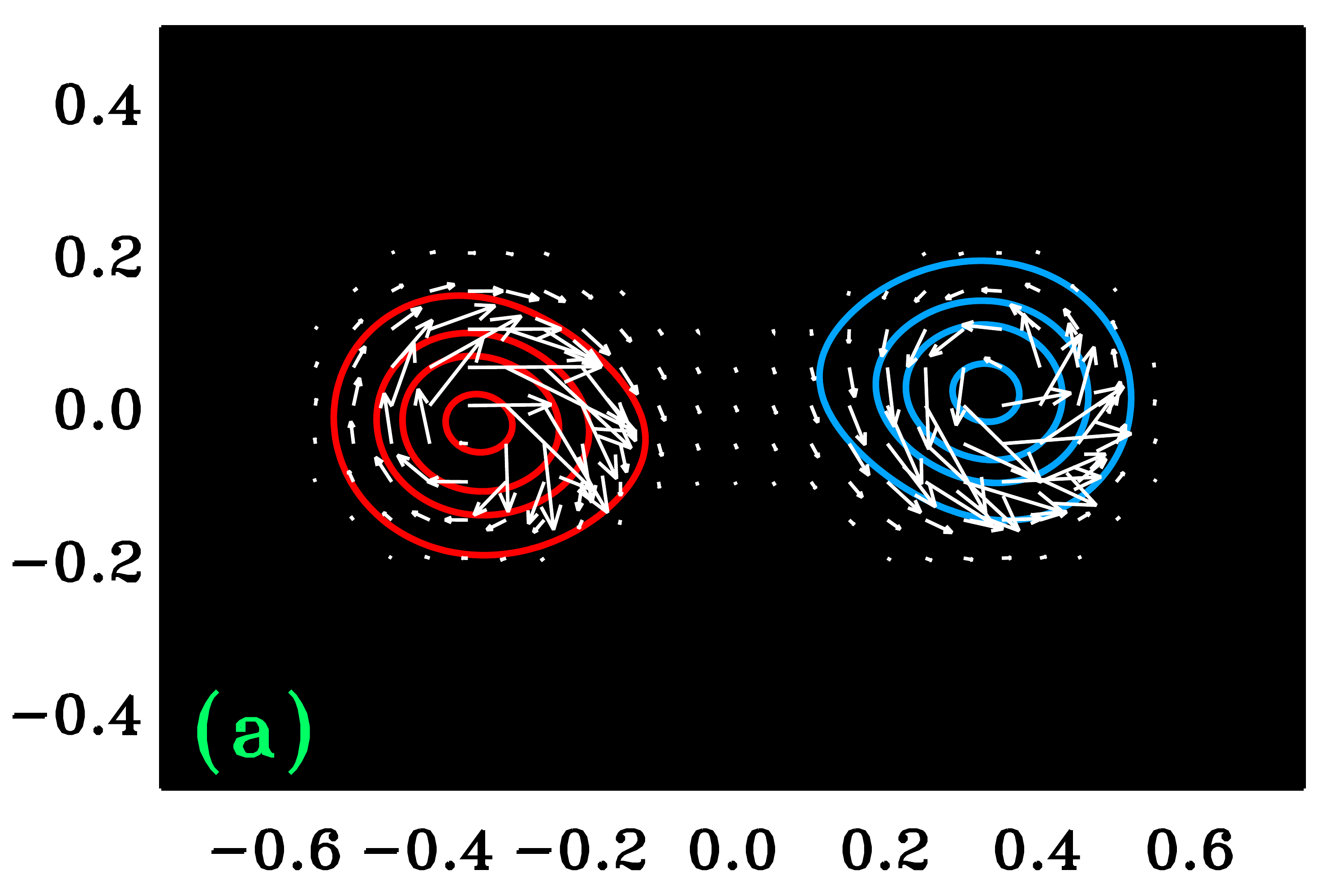}\hfil\includegraphics[width=0.32\textwidth]{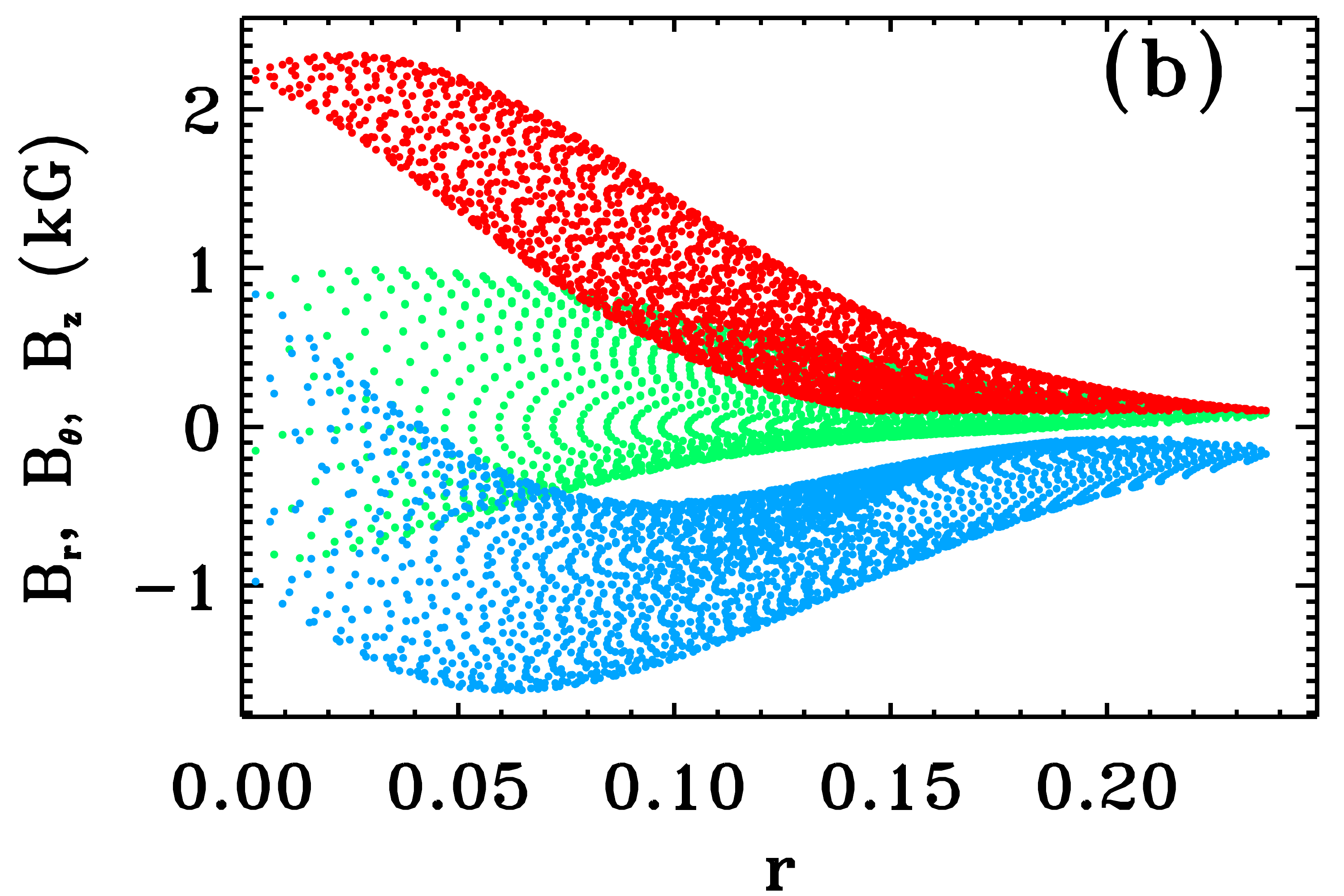}\hfil\includegraphics[width=0.32\textwidth]{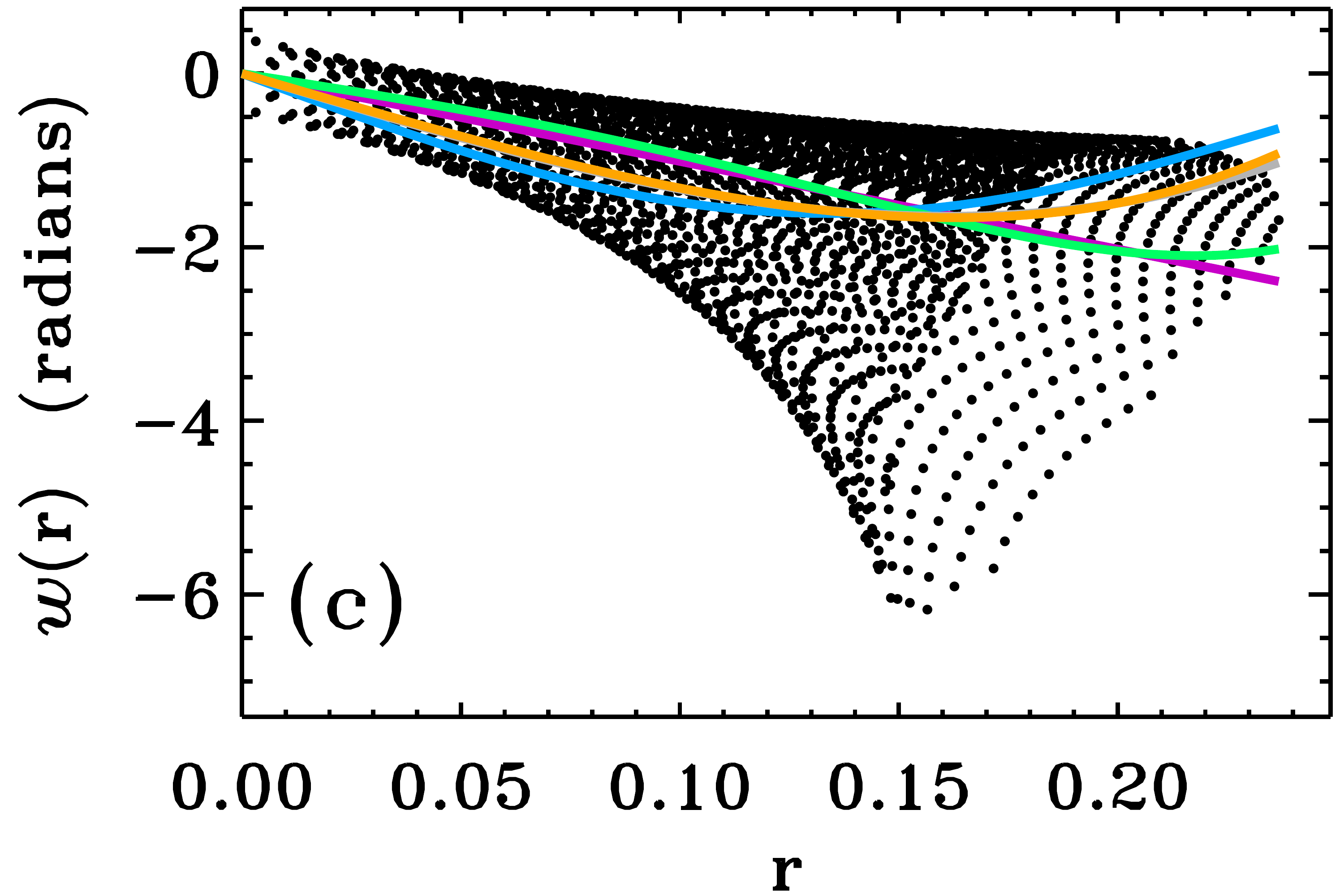}
\\[2mm]
\includegraphics[width=0.32\textwidth]{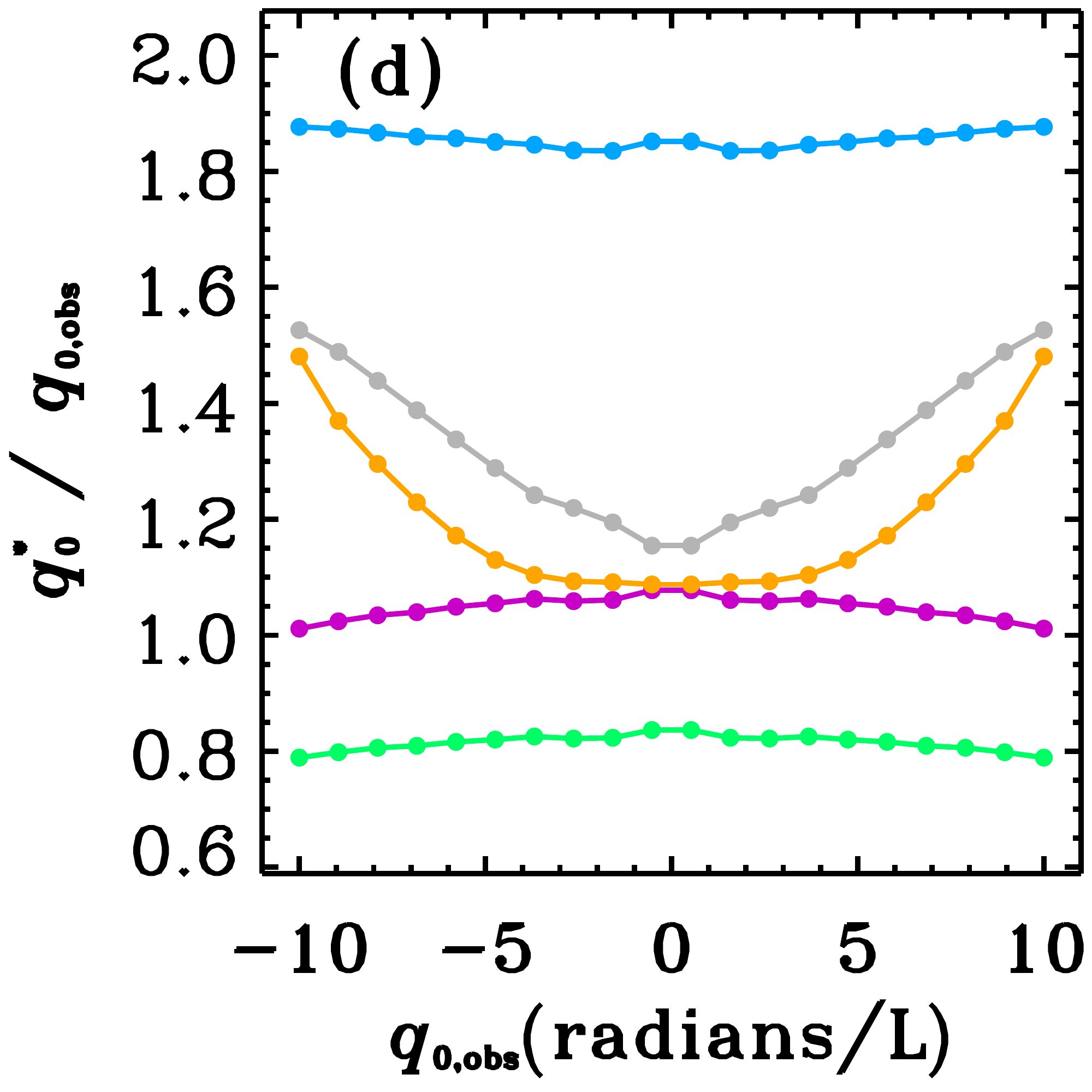}\hfil\includegraphics[width=0.32\textwidth]{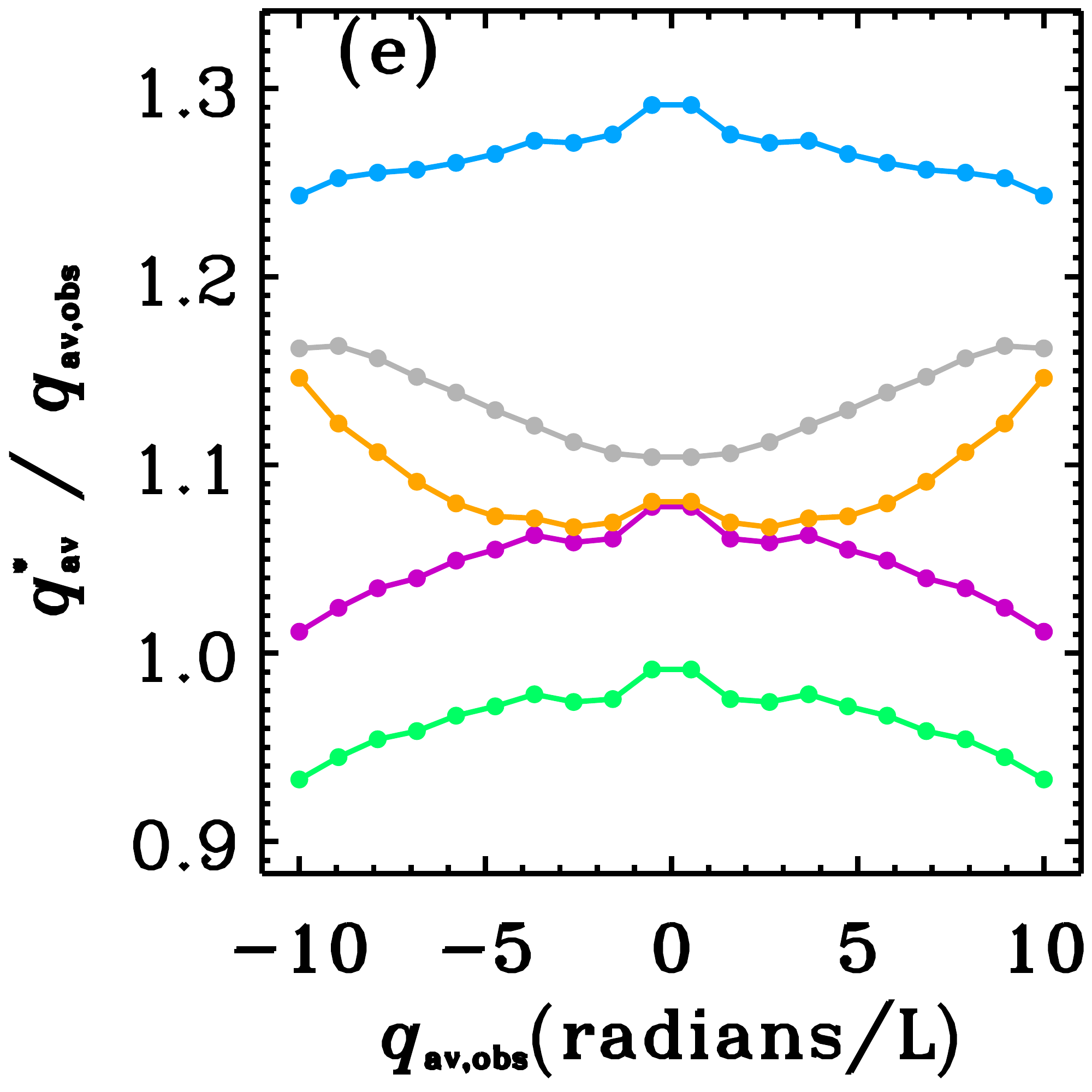}\hfil\includegraphics[width=0.32\textwidth]{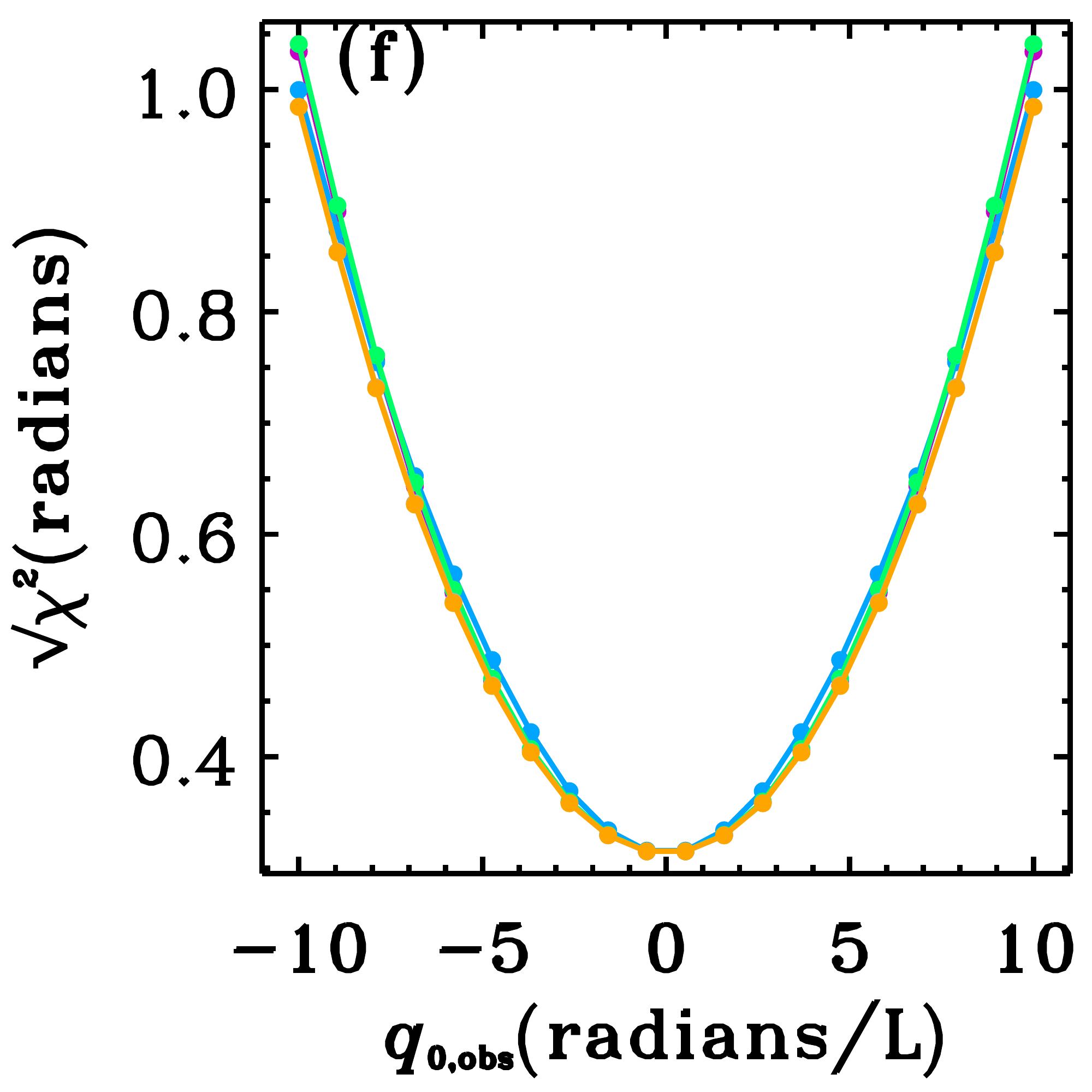}
\caption{Same as Figure~\ref{fgz0}, except that the magnetic field is sampled on  the plane $z_0= 0.14375 \, \fgl$.
(a) Vector magnetic field at $z_0= 0.14375 \, \fgl$ for the toroidal flux loop  with $(\ql/\al) =-10 $~radians $ \fgl^{-1}$.
The maximum horizontal field strength shown is approximately 1730~G.
(b) Inferred cylindrical components of the magnetic field [$B_r$, $B_\theta$, $B_z$] as a function of radial distance from the flux-loop axis [$r$], for the positive-polarity footpoint of the  magnetic field shown in (a).
(c) The ratio [$\wrat = B_\theta / B_z$] as a function of radial distance from the flux-loop axis [$r$].
(d) The ratio of the inferred and true winding rates at the flux-tube axis  [$\wind_0^\ast / \wind_{0,{\rm obs}}$] as a function of the true winding rate at the axis [$\wind_{0,{\rm obs}}$] for the fitting models in (c).
(e) Same as (d) except for the corresponding ratio of the inferred and true average winding rates [$\wind_{\rm av}^\ast / \wind_{\rm av,obs}$] as a function of the true average winding rate [$\wind_{\rm av,obs}$], with the average taken over the range used for the fitting.
(f) The corresponding value of $\sqrt{ \chi^2 }$ (see Equation~(\ref{chisq})) as a function of $\wind_{0,{\rm obs}}$.
}
\label{fgz2}
\end{figure}

Figure~\ref{fgz2} shows results for the case where the magnetic field is sampled on the plane $z_0= 0.14375 \, \fgl$.
Compared to the case shown in Figure~\ref{fgz0}, this value of $z_0$ may be taken to represent the flux loop at an earlier stage of emergence (\myeg{} \citeauthor{fan03} \citeyear{fan03}, \citeyear{fan04}).
On the plane $z_0= 0.14375 \, \fgl$ several of the fitting-model assumptions are violated, these include:
the axis of the flux loop is curved  and not parallel to the $z$-direction,
the magnetic field varies in the $z$-direction (although the variation is quite different from that discussed in Appendix~\ref{sec_alpha}),
the radial component of the magnetic field is nonzero and not single-valued (see Figure~\ref{fgz2}(b)), 
and the ratio $\wrat = B_\theta / B_z$ is not single-valued for a fixed value of $r$ (Figure~\ref{fgz2}(c)).
The latter two features are observable and indicate that the fitting model is not strictly appropriate for this data set.
However, judging from the vertical component of the field $B_z$ in Figure~\ref{fgz2}(a), the contour for $B_z=100$~G is fairly circular and it may seem reasonable to apply the fitting methods without modification.
Taking this approach, we find that none of the fitting methods retrieve the correct values for the on-axis or average winding rates (Figure~\ref{fgz2}(d) and (e)), although in some cases the estimates are within approximately 10\,\% of the true values, such as for the fitting model with a uniform winding rate.
We also find  cases where the estimates retrieved by two different fitting models approximately agree, yet both estimates do not agree with the true winding rate.
The quality of the fits  is universally poor (Figure~\ref{fgz2}(f)); that is, according to  $\chi^2$ the various best-fit models are effectively indistinguishable, yet they produce a variety of winding-rate estimates.
For these test cases we find that the inferred winding rates are not directly proportional to true values.
We also find that slightly better estimates for both the on-axis and average winding rates (with the average taken over the fitting interval) and smaller $\chi^2$ values can be retrieved by restricting the fitting to observation points closer to the axis, where the scatter in $\wrat_{\rm obs}$ is less severe, but this is not true for all values of $z_0$.

\begin{figure}[ht]
\includegraphics[width=0.32\textwidth]{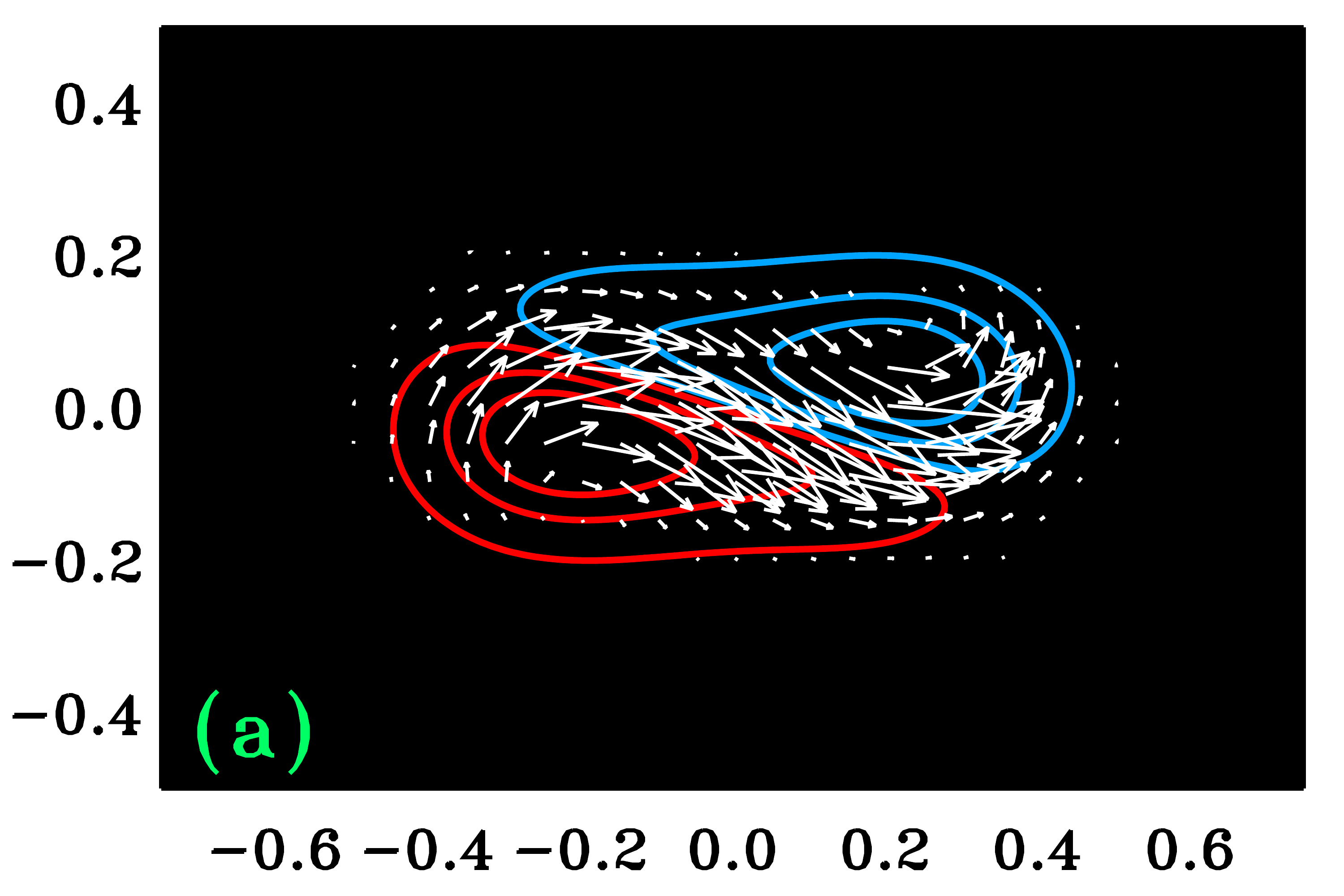}\hfil\includegraphics[width=0.32\textwidth]{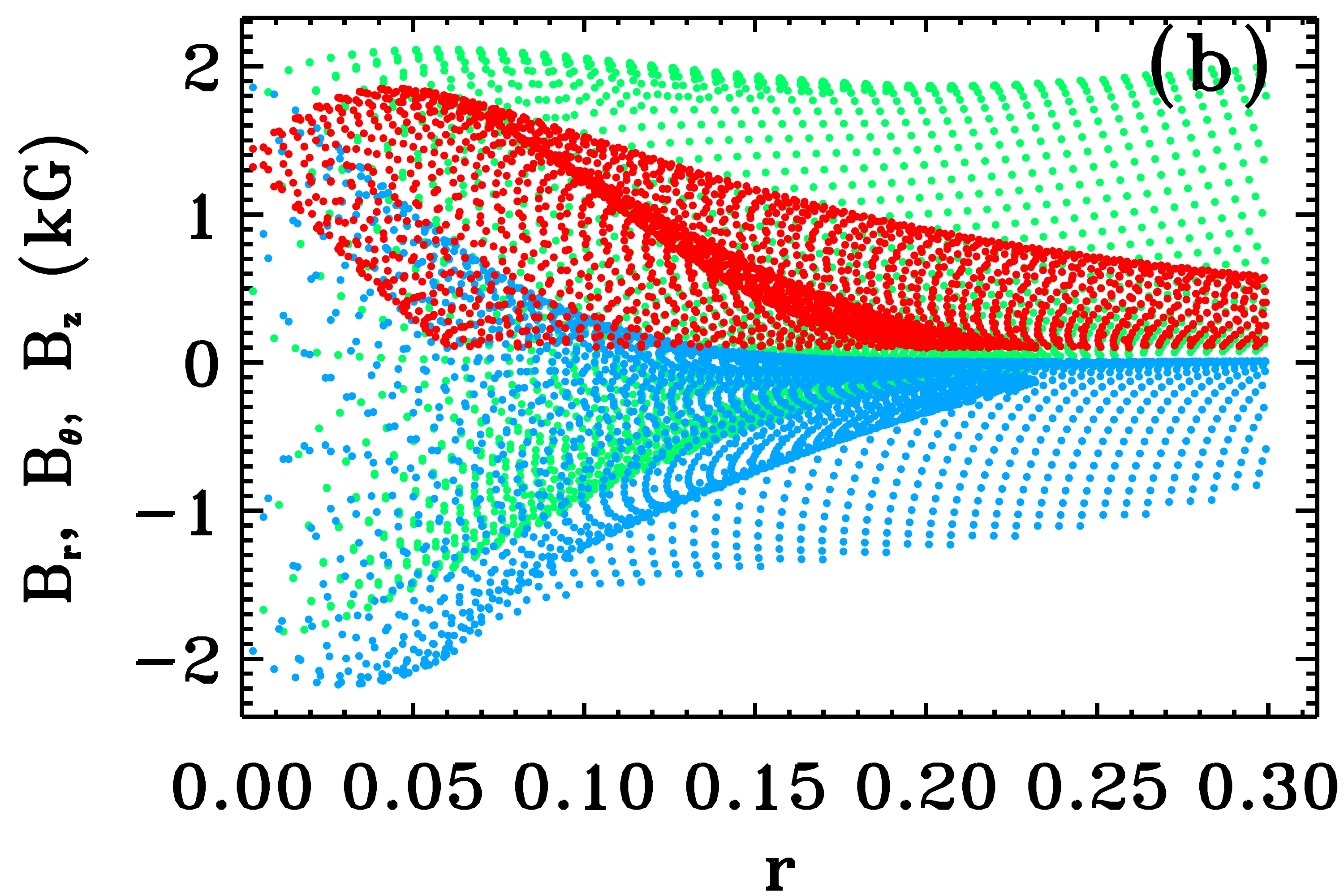}\hfil\includegraphics[width=0.32\textwidth]{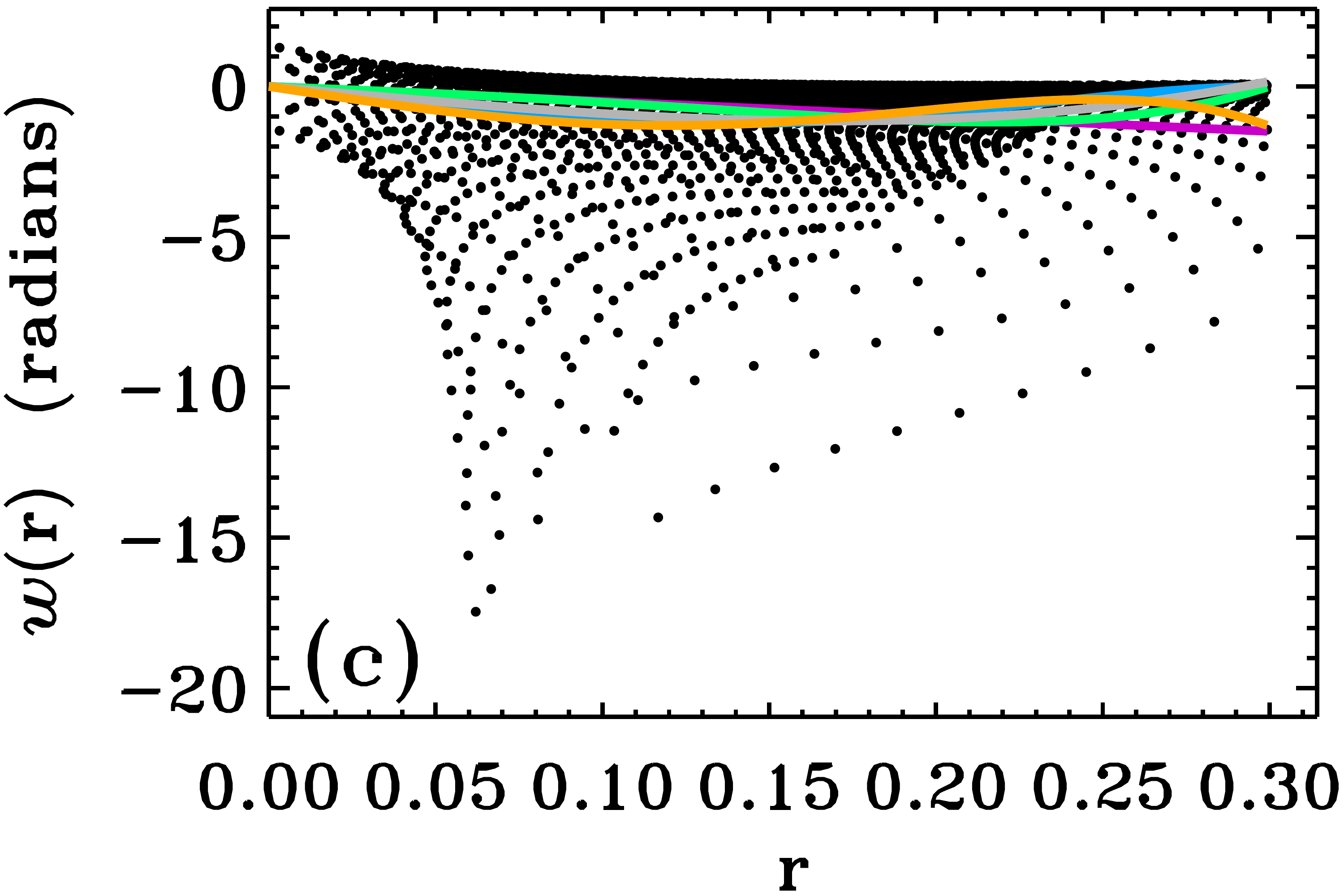}
\\[2mm]
\includegraphics[width=0.32\textwidth]{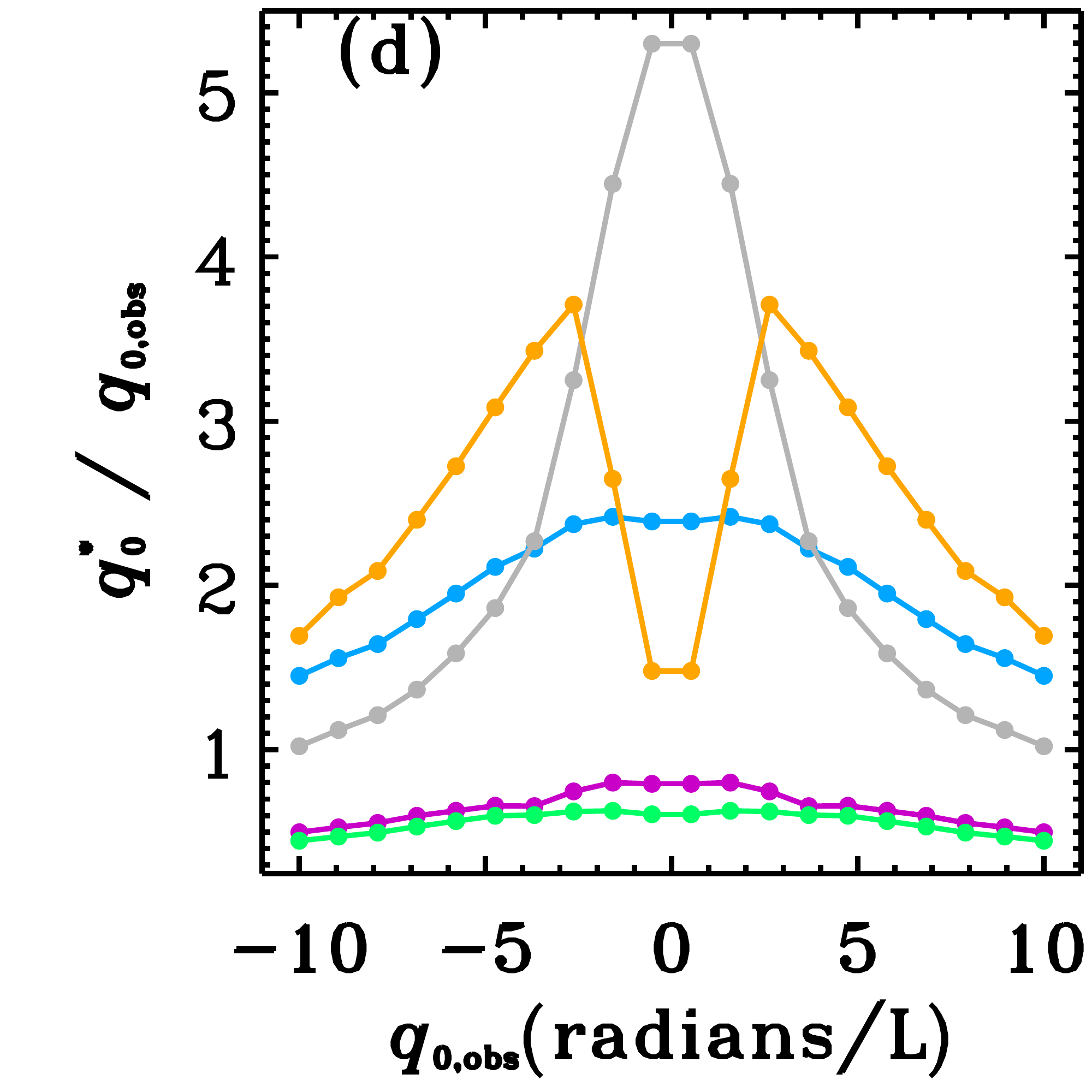}\hfil\includegraphics[width=0.32\textwidth]{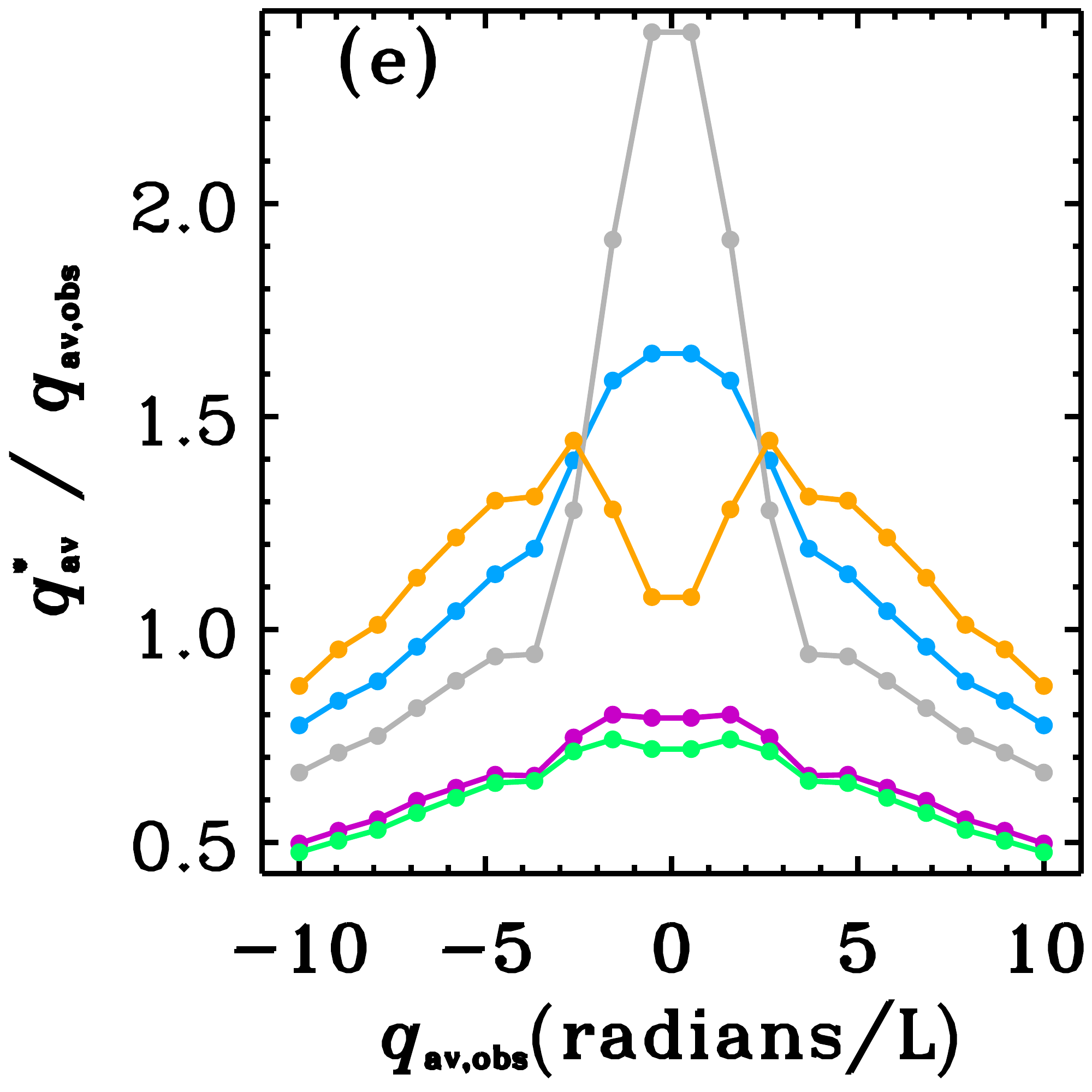}\hfil\includegraphics[width=0.32\textwidth]{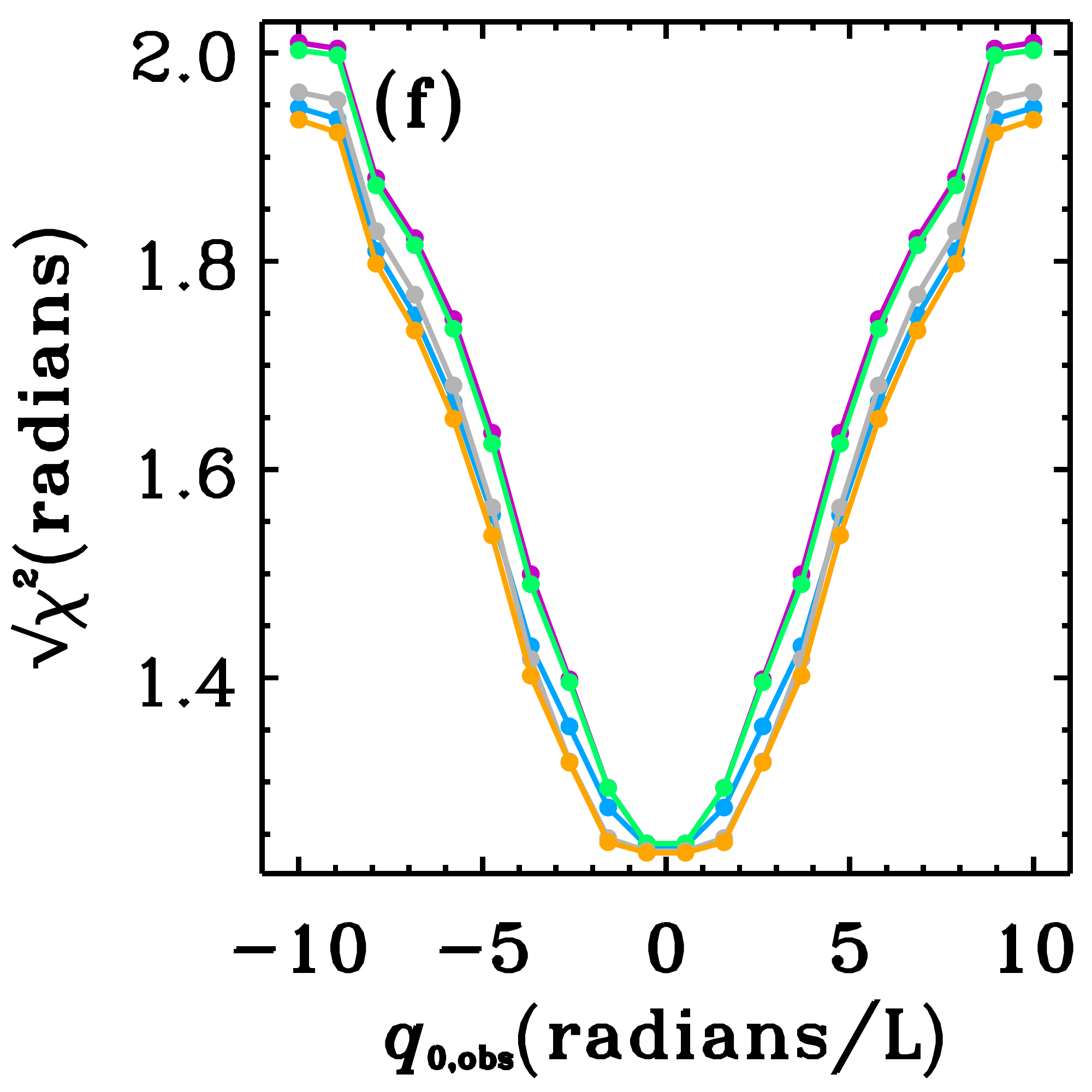}
\caption{Same as Figure~\ref{fgz0}, except that the magnetic field is sampled on  the plane $z_0= 0.306250 \, \fgl$.
(a) Vector magnetic field at $z_0= 0.306250 \, \fgl$ for the toroidal flux loop  with $(\ql/\al) =-10 $~radians $ \fgl^{-1}$.
The maximum horizontal field strength shown is approximately 2312~G.
(b) Inferred cylindrical components of the magnetic field  [$B_r$, $B_\theta$, $B_z$] as a function of radial distance from the flux-loop axis [$r$], for the positive-polarity footpoint of the  magnetic field shown in (a).
(c) The ratio [$\wrat = B_\theta / B_z$] as a function of radial distance from the flux-loop axis [$r$].
(d) The ratio of the inferred and true winding rates at the flux-tube axis [$\wind_0^\ast / \wind_{0,{\rm obs}}$] as a function of the true winding rate at the axis [$\wind_{0,{\rm obs}}$] for the fitting models in (c).
(e) Same as (d) except for the corresponding ratio of the inferred and true average winding rates [$\wind_{\rm av}^\ast / \wind_{\rm av,obs}$] as a function of the true average winding rate [$\wind_{\rm av,obs}$], with the average taken over the range used for the fitting.
(f) The corresponding value of $\sqrt{ \chi^2 }$ (see Equation~(\ref{chisq})) as a function of $\wind_{0,{\rm obs}}$.
}
\label{fgz4}
\end{figure}

In other experiments we find that  the performance of the fitting methods gets progressively worse as $z_0$ increases. 
Figure~\ref{fgz4} shows the case where the magnetic field is sampled on the plane $z_0= 0.306250 \, \fgl$. 
In this case, the footpoints are not well separated, the magnetic field is obviously not axisymmetric, the flux-loop axis is highly inclined with respect to the $z$-axis, and the winding rate clearly has a large apparent azimuthal dependence.
None of the fitting models consistently retrieve an accurate estimate for the winding rates.
For larger values of the true winding rates, the accuracy of the winding-rate estimates retrieved by the best-fit polynomial basis-function approximations gets worse as $n_b$ increases, contrary to expectations (when the magnetic field does satisfy the assumptions built into the fitting models).
These results indicate that the fitting methods that infer the radial variation of the winding rate do not generally retrieve reliable winding-rate estimates when the observed magnetic-flux tube deviates substantially from the fitting model.

\section{Conclusions}
\label{sec_conc}

We demonstrate how assumptions built into models for twisted magnetic-flux tubes used in least-squares fitting methods can influence estimates retrieved for the winding rate of the magnetic-field lines about the flux-tube axis.
The fitting methods that we examine estimate the winding rate in a magnetic-flux tube by finding the value of a parameter (or set of parameters), related to the winding rate, that corresponds to the minimum of the discrepancy between observations and predictions from a flux-tube model, in a similar fashion to \inlinecite{Nandyetal2008}.
For the flux-tube model used in the fitting we assume that the magnetic field is static, axisymmetric, and does not vary in the vertical direction.
Using error-free, synthetic vector magnetic-field data constructed with models for twisted magnetic-flux tubes we test the performance of these fitting methods at recovering the true on-axis and average winding rates.
We find that the accuracy of the winding-rate estimates retrieved with fitting methods is  sensitive to the details of the assumptions built into the flux-tube model used for the fitting.

We identify the radial variation of the winding rate  as one assumption that can have a significant impact on the winding-rate estimates.
For fitting methods that make a fixed assumption about the radial variation of the winding rate, we find that a significant error can be introduced into the winding-rate estimates if an incorrect assumption is made about the radial dependence of the winding rate.
We show mathematically that the magnitude of the discrepancy between the inferred and the true winding rates is related to the mismatch between the radial variation of the winding rate assumed for the  flux-tube model used in the fitting and that used to construct the synthetic data.
We subsequently show that fitting methods that make different assumptions about the radial variation of the winding rate can produce different estimates for the winding rates when applied to the same synthetic data.
We  find that best-fit models with smaller values of the misfit function [$\chi^2$] do not necessarily  retrieve more accurate winding-rate estimates.
We also find that a model that assumes a uniform winding rate within the flux tube does not necessarily retrieve an accurate estimate for the radial average of the winding rate, as one may expect.
We show that the errors caused by making an incorrect assumption about the radial dependence of the winding rate can be largely avoided by inferring the radial variation of the winding rate, using a basis-function approximation (\myeg{} \opencite{1992nrfa.book.....P}).

We apply the fitting methods  to synthetic data generated with a twisted, toroidal, magnetic-flux-loop model (\myeg{} \citeauthor{fan03}, \citeyear{fan03}, \citeyear{fan04})  which has some features that are more solar-like than the other synthetic data sets  used in this investigation.
We find that the correct winding rates can be recovered by fitting methods if the magnetic-flux loop is sampled on a special plane where the flux-loop axis is parallel to the vertical direction, and the important assumptions built into the  model used for the fitting are satisfied.
If the magnetic-flux loop is sampled away from this special plane the flux loop can differ substantially  from the  model used for the fitting, in ways that are expected for solar observations of twisted magnetic-flux tubes.
In these cases we find that fitting methods generally fail to recover the correct winding rates, with discrepancies between the inferred and true winding rates of up to several tens of percent.

We therefore conclude that the models used for the fitting, based on static, axisymmetric, vertically translation-invariant flux tubes, are too simple to yield accurate estimates for the winding rate of the magnetic field in solar magnetic structures in general.
However, if the observed magnetic field is not substantially different from the simple model used for the fitting (or  information is available to justify the choice of fitting model) these methods may be useful for providing a rough estimate of the winding rate for the plane where the magnetic field is measured, 
provided that fitting methods that infer the radial variation of the winding rate are used.
Beyond rough estimates,  however, a more sophisticated model for twisted magnetic-flux tubes is required for least-squares fitting methods to retrieve accurate winding-rate estimates in general.

\begin{acks}
The author thanks Graham Barnes for helpful comments and discussion.
The author thanks the referee for constructive criticism.
This material is based upon work supported by the National Science Foundation under Grants No.~0454610 and 0519107.
\end{acks}

\appendix

\section{Testing the Fitting Methods with Twisted Magnetic Fields that Vary in the Vertical Direction}
\label{sec_alpha}

To show how the fitting methods perform when the observed flux tube has a magnetic field that varies in the $z$-direction, we consider an axisymmetric, linear force-free, magnetic field (\myeg{} \opencite{1965IAUS...22..337S}; \opencite{1983ApJ...266..848B}; \opencite{2008ApJ...681.1660P}), governed by the magnetic-flux function

\begin{equation}
\Psi (r, z)  = \Psi_0 r  J_1 ( k r ) \exp ( - m  z ) /  ( R  J_1 ( k R )) \, ,
\label{alphaeqn}
\end{equation}

\noindent
where
$J$ is the Bessel function of the first kind,
$m$ is a parameter,
$k^2=m^2+\alpha^2$,
and $\alpha$ is the (constant) force-free parameter.
We assume that $m$ is real and positive so the magnetic-flux function described by Equation~(\ref{alphaeqn}) decays exponentially with increasing height $z$.
The magnetic-field lines lie in the surfaces of constant $\Psi$.  
We assume that the external boundary of the flux tube is made up of the field lines on the flux surface $\Psi=\Psi_0$, and $R$ is the radius of the flux tube at $z=0$ [$\Psi (R, 0)  = \Psi_0$].
The magnetic-field components are:

\begin{eqnarray}
B_r (r, z)     & =  &  - \frac{1}{r} \frac{\partial \Psi}{\partial z} =  \Psi_0  m  J_1 ( k r ) \exp ( - m  z ) / ( R  J_1 ( k R ))  \, ,  \label{alphabr}\\
B_\theta (r, z) & = & \frac{\alpha}{r} \Psi = \Psi_0  \alpha  J_1 ( k r ) \exp ( - m  z ) / ( R  J_1 ( k R ))  \, , \label{alphabt}\\
B_z (r , z)    & = &  \frac{1}{r} \frac{\partial \Psi}{\partial r} = \Psi_0   k J_0 ( k r ) \exp ( - m z ) / ( R  J_1 ( k R ))  \, . \label{alphabz}
\end{eqnarray}

\noindent
This magnetic field is useful for constructing synthetic data to test the fitting methods because it has some features that are consistent with the fitting model (\myie{} the magnetic field is axisymmetric, and the flux-tube axis is straight and vertical). 
This allows us to demonstrate the influence of the various features of this magnetic field that are not included in the fitting model, which are:
the magnetic field varies in both the vertical and radial directions,
the radial component of the magnetic field is non-zero,
and the radius of the flux tube varies in the vertical direction (\myie{} the radial location of the flux surface with $\Psi=\Psi_0$ varies in the $z$-direction).

For $\alpha \neq 0$ this magnetic field is twisted (\myie{} $B_\theta \neq 0$, see Equation~(\ref{alphabt})).
At a given position, the winding rate of a field line about the flux-tube axis, per unit length along the axis, is

\begin{equation}
\wind_{_{\rm LFF}}  = \frac{B_\theta }{  r B_z } = \frac{\alpha J_1 ( k r )}{ k r J_0 ( k r ) } \, ,
\label{alphawind}
\end{equation}

\noindent
which varies in the radial direction but not in the vertical direction.
At the flux-tube axis ($r=0$), the winding rate is $\wind_{0,{\rm obs}} = \lim_{r \to 0} \wind_{_{\rm LFF}}  = \alpha / 2$.

As the winding rate varies only in the radial direction, we can test the various fitting methods with error-free synthetic data generated with this magnetic field using the same approach used in Sections~\ref{sec_sing} and \ref{sec_multi}.
To this end, in Equation~(\ref{chisq}) we set $\wrat_{{\rm obs},i} = r_i \wind_{_{\rm LFF}} (r_i)$.
For synthetic data generated with this magnetic field it can be shown that the best-fit on-axis and average winding rates are not directly proportional to the respective true values.
Therefore, we test the performance of the fitting methods over a range of winding rates, choosing a typical range of values as follows.
For a flux loop of length $\myl$, the critical twist typically quoted for the onset of the kink instability is $| \wind \myl | \approx 2 \pi$~radians (\myeg{} \opencite{1972SoPh...22..425R}; \opencite{HoodPriest1979}; \opencite{1983SoPh...88..163E}; \opencite{1990ApJ...361..690M}; \opencite{Vellietal1990}; \opencite{1998ApJ...494..840L}; \opencite{vanderLindenHood1998}, \citeyear{vanderLindenHood1999}; \opencite{1998A+A...333..313B}; \opencite{Baty2001}; \opencite{fan03}, \citeyear{fan04}; \opencite{tor04}).
For the magnetic-flux tube described by Equation~(\ref{alphaeqn}), we consider a vertical section of length $ \delta \myl = 4$~Mm, taken to represent part of a larger magnetic-flux system.
We construct twenty synthetic data sets with twenty equally spaced values for $\wind_{0,{\rm obs}}$ that yield partial twist values at the axis in the range $ - \pi / 2  \le \wind_{0,{\rm obs}} \delta \myl \le \pi / 2$; the case where the winding rate at the axis is zero is not considered.
We set $m= 0.08~\rm{Mm}^{-1}$ so that the magnetic-flux function decreases by approximately 27\,\% from $z=0$ to $z= \delta \myl =4$~Mm (at a fixed $r$).
For this range of parameter values it can be confirmed that $B_z$ is single signed in the volume bounded by the flux surface $\Psi=\Psi_0$ and  $0 \le z \le \delta \myl$; we have restricted our attention to this volume to avoid locations where the winding rate for this field is infinite (corresponding to the zeroes of the Bessel function $J_0$, \myie{} $B_z=0$).
Synthetic measurements are generated at twenty equally-spaced observation points in the interval $0 \le r_i \le  R$, all of which are used for the fitting, the average winding rate is calculated for the interval $[ 0 , R ]$, and we set $R=2$~Mm.
For this choice of $R$ the spacing between observation points (in the radial direction) is approximately 100~km, consistent with the spatial resolution provided by data from instruments such as the {\it Solar Optical Telescope} onboard {\it Hinode} (\myeg{} \opencite{hinode}; \opencite{2008SoPh..249..167T}).

\begin{figure}[ht]
\begin{center}
\includegraphics[width=0.32\textwidth]{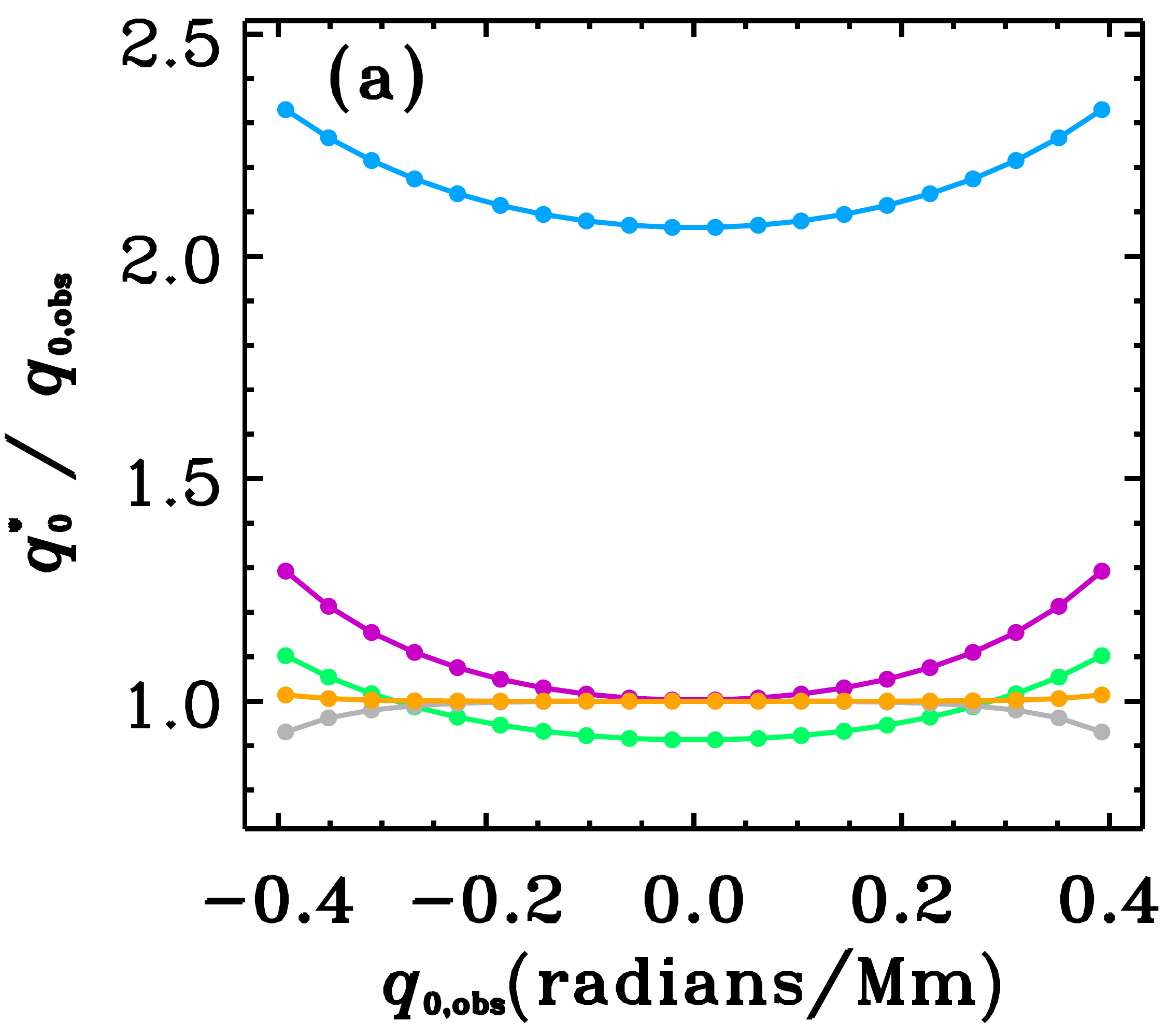}\hfil\includegraphics[width=0.32\textwidth]{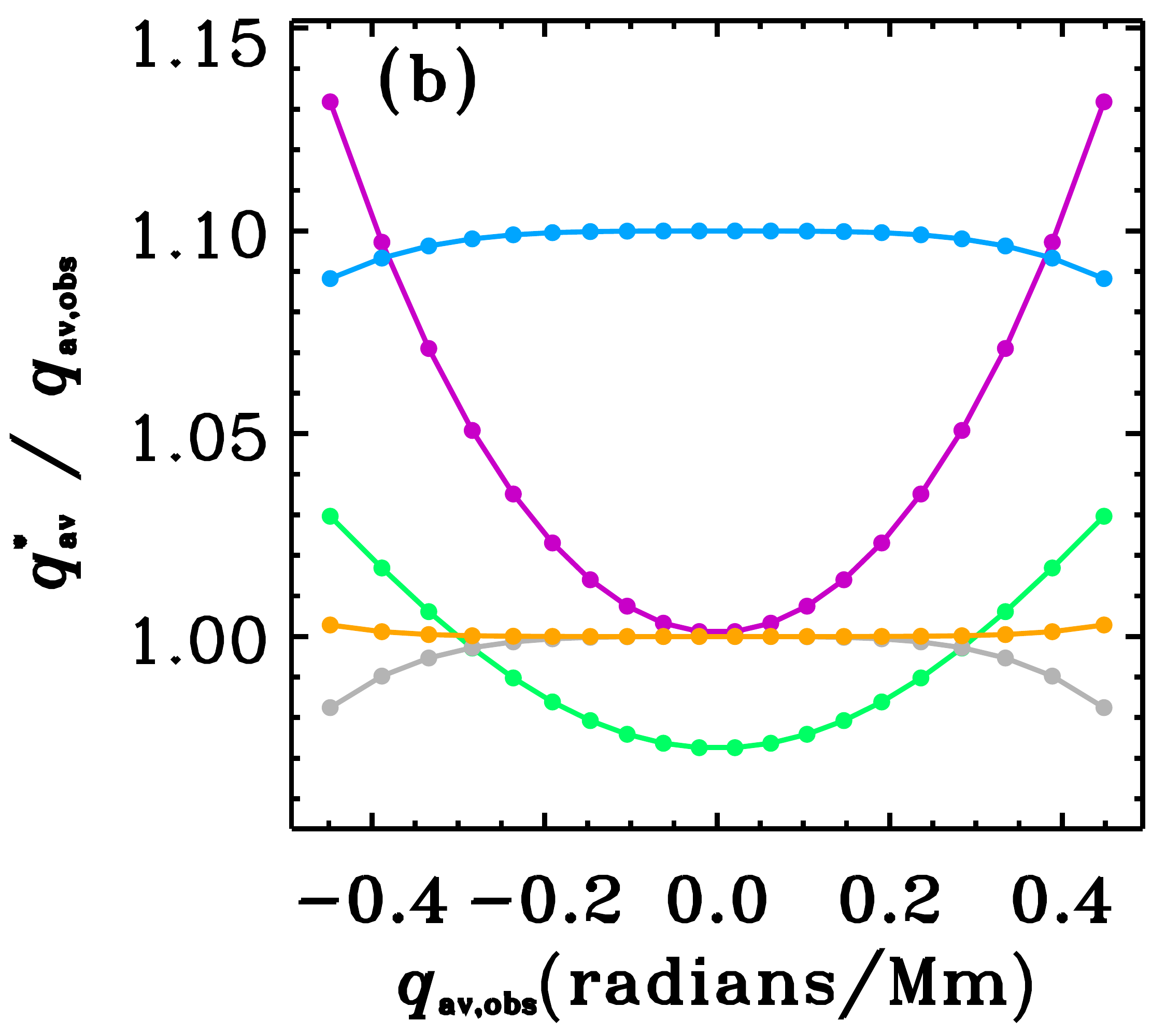}\hfil\includegraphics[width=0.32\textwidth]{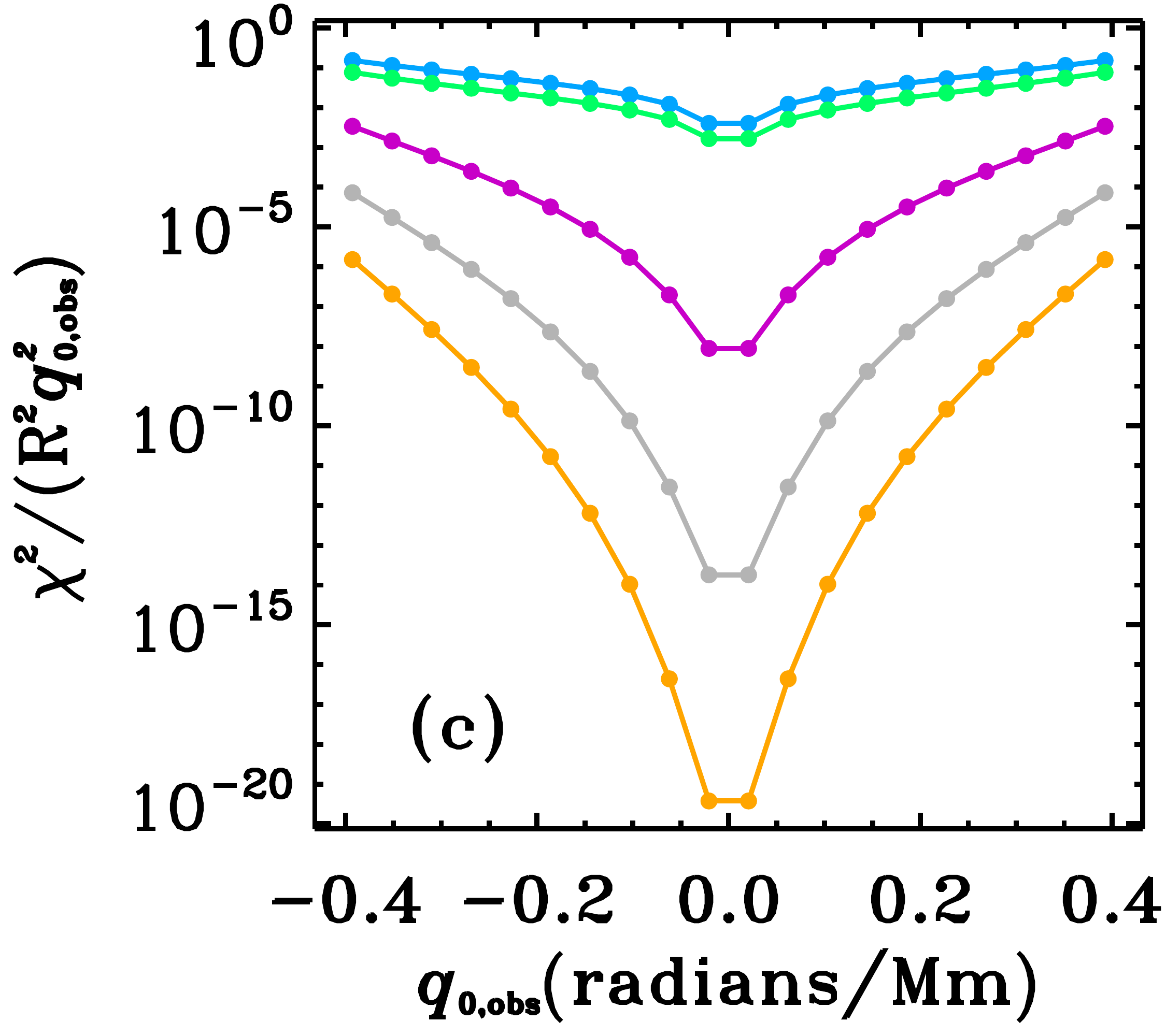}
\end{center}
\caption{
(a)
The ratio of the inferred and true winding rates at the flux-tube axis [$\wind_0^\ast / \wind_{0,{\rm obs}}$] as a function of the true winding rate at the axis [$\wind_{0,{\rm obs}}$] for different fitting models.
In these tests we use the linear force-free magnetic field (see Equation~(\ref{alphawind})) to generate twenty synthetic data sets with a range of winding rates $\wind_{0,{\rm obs}}$ with fixed $m$ (see text).
Each point represents the result from a single fitting experiment; the curves joining the points are only included as a guide.
The purple curve is the best-fit case for a fitting model with $\qmod (r) = \qmodamp$,
the blue curve is for $ \qmod (r) = \qmodamp (1-(r/R)^2)^2$,
the green curve is for $\qmod (r) = \qmodamp ( 1 + 2 (r/R)^2 - 3 (r/R)^4 )$,
the gray curve is for $\qmod (r) =  \qbamp_1  + \qbamp_2 (r/R)^2$,
and the orange curve is for $\qmod (r) =  \qbamp_1  + \qbamp_2 (r/R)^2 + \qbamp_3 (r/R)^4$.
(b)
Same as (a) except for the corresponding ratio of the inferred and true average winding rates [$\wind_{\rm av}^\ast / \wind_{\rm av,obs}$] as a function of the true average winding rate [$\wind_{\rm av,obs}$], with the average taken over the range used for the fitting.
(c)
The corresponding value of $\chi^2 / (R^2 \wind_{0,{\rm obs}}^2)$ (see Equation~(\ref{chisq})) for the best-fit model as a function of $\wind_{0,{\rm obs}}$.
}
\label{figvarz}
\end{figure}

\begin{figure}[ht]
\begin{center}
\includegraphics[width=0.49\textwidth]{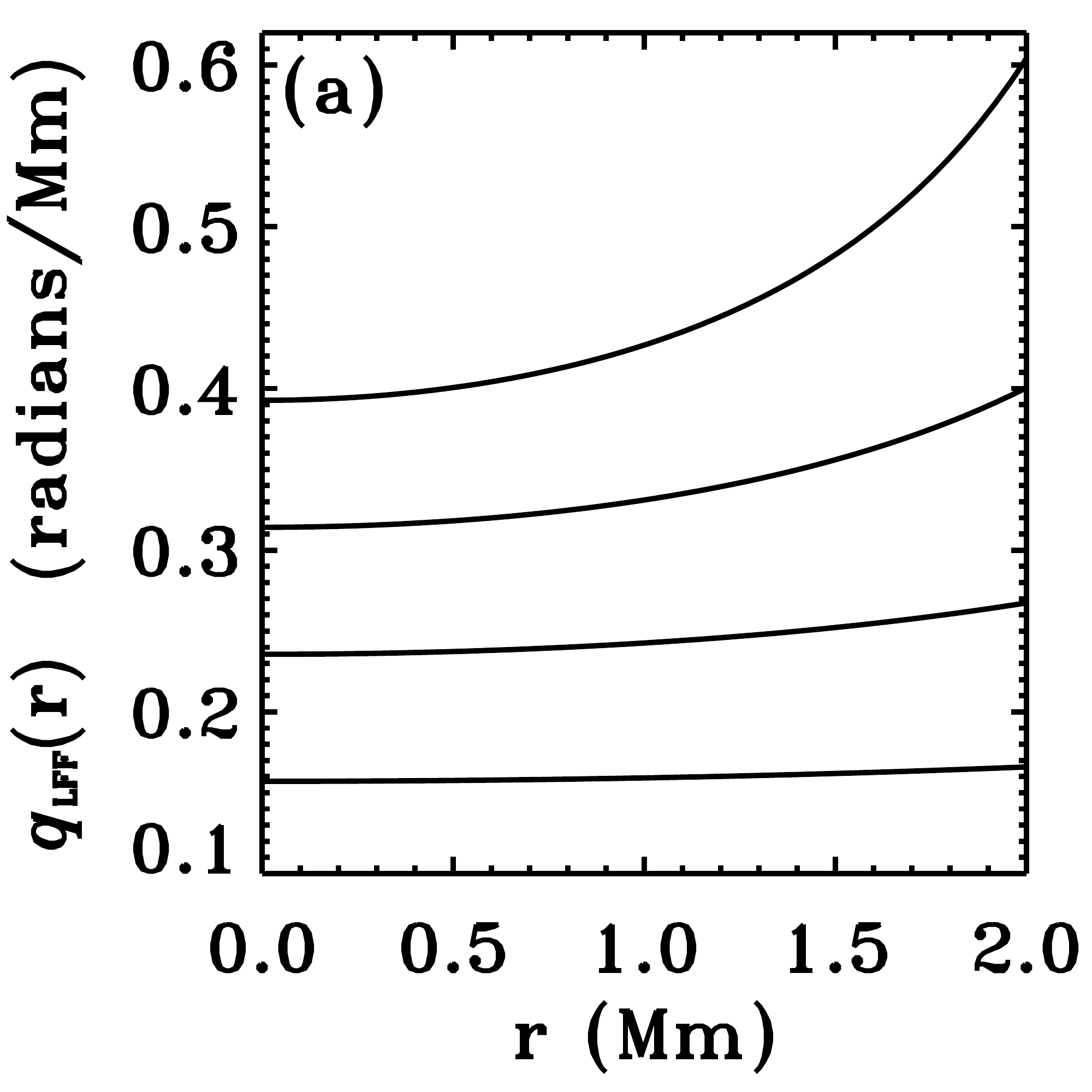}\hfil\includegraphics[width=0.49\textwidth]{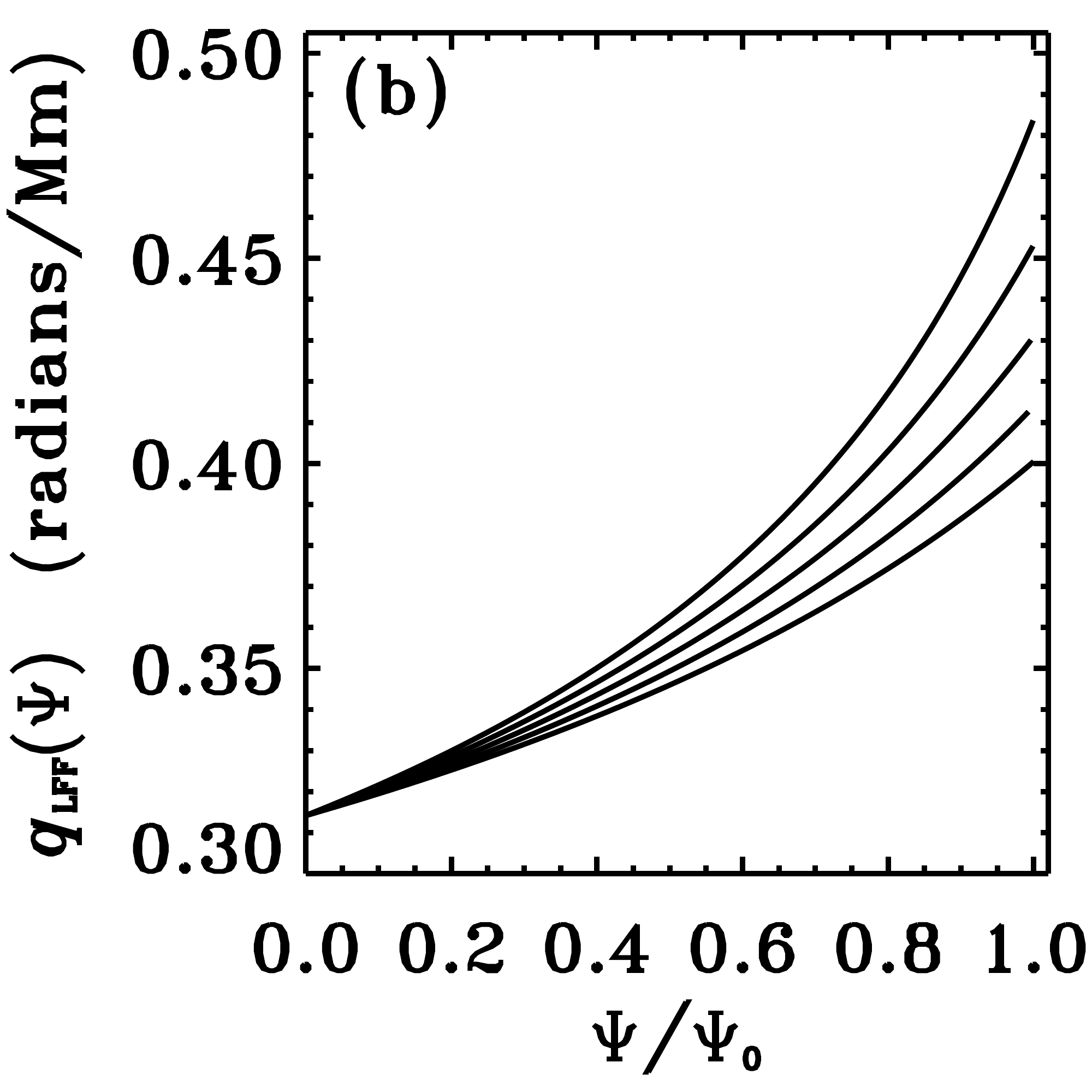}
\end{center}
\caption{(a) Winding rate [$\wind_{_{\rm LFF}}$] as a function of $r$ (see Equation~(\ref{alphawind})) for a flux tube with $m= 0.08~\rm{Mm}^{-1}$.
The different curves correspond to four different values for the winding rate at the flux-tube axis: $\wind_{0,{\rm obs}} = \alpha / 2 = \{ 0.05, 0.075, 0.1, 0.125 \} \pi$~radians~Mm$^{-1}$.
(b) Winding rate [$\wind_{_{\rm LFF}}$] as a function of $\Psi / \Psi_0$ for a flux tube with $m= 0.08~\rm{Mm}^{-1}$ and $\alpha = 0.2 \pi$~radians~Mm$^{-1}$.
The different curves correspond to five different heights: $z= {0, 1, 2, 3, 4}$~Mm, with greater heights corresponding to curves with larger values for $\wind_{_{\rm LFF}}$.
}
\label{qlff}
\end{figure}

The results of this exercise (see Figure~\ref{figvarz}) are qualitatively similar to those discussed in Section~\ref{sec_multi}.
For example, we find that both the on-axis and average winding rates retrieved by the best-fit polynomial basis-function approximations are generally more accurate (and the $\chi^2$ values are smaller) than those retrieved with fitting models with a fixed radial variation for the winding rate; although for the case shown in Figure~\ref{figvarz} the magnitude of the discrepancy between the inferred and true winding rates is generally smaller than in Tables~\ref{demo2tab} and \ref{toytab}.
We also find that the various fitting models which make  fixed assumptions about the radial variation of the winding rate can produce  different estimates for the winding rates given the same synthetic data, as in Table~\ref{toytab}.
Several additional points are worth noting:
i) For the best-fit polynomial basis-function approximations the magnitude of the discrepancy tends to increase with increasing magnitude of the winding rate at a fixed value for $m$; the same is also true for the fitting model that assumes the winding rate is constant.
This can be understood by referring to Figure~\ref{qlff}(a) which shows how the winding-rate profiles (for a fixed value for $m$) vary with $\wind_{0,{\rm obs}}$. 
For small values of $\wind_{0,{\rm obs}}$ the winding-rate profile is almost constant over the flux-tube interior, but as $\wind_{0,{\rm obs}}$ increases the change in the winding rate from the axis to the external boundary increases.
This is because higher order terms in Taylor-series expansion for $\wind_{_{\rm LFF}} (r)$ about $r=0$ become more important as $\wind_{0,{\rm obs}}$ (or $\alpha$) increases.
Consequently, low-order polynomial basis-function approximations (and the fitting model with a uniform winding rate) are less accurate at large values of $\wind_{0,{\rm obs}}$.
ii) In other experiments with different values for $m$ but the same range of values for $\wind_{0,{\rm obs}}$ we find that the magnitude of the discrepancy between the inferred and true winding rates tends to increase slightly as $m$ increases.

If the observed magnetic field varies in the $z$-direction, the results in Figure~\ref{figvarz} indicate that reasonable estimates for the winding rates can be obtained by fitting methods that infer the radial variation of the winding rate, for the plane where the magnetic field is measured, provided that the observed field is axisymmetric and has a flux-tube axis is straight and vertical.
However, because the magnetic field varies in the $z$-direction the following considerations must be emphasised regarding the validity of the winding-rate estimates away from the plane where the magnetic field is measured.

For an axisymmetric magnetic-flux tube, with an axis that is straight and vertical, and with a magnetic field that may vary in the $z$-direction, the winding rate at the flux-tube axis does not vary in the $z$-direction (see, \myeg, \opencite{1989GApFD..48..217F}, their Equation~(4.15)).
Thus, if an accurate estimate for the winding rate at the flux-tube axis can be retrieved at one height (by any method), it can be used to estimate the winding rate at the axis at other heights, provided that the observed flux tube is indeed axisymmetric with an axis that is straight and vertical.

Away from the flux-tube axis, an axisymmetric magnetic-flux tube can generally have a winding rate that varies in the $z$-direction (see, \myeg, \opencite{1983ApJ...266..848B}, their Equation~(5.7)).
Therefore, estimates for the winding rate away from the flux-tube axis, and thus the average winding rate, retrieved with fitting methods are generally valid only for the plane where the magnetic field is measured.
The magnetic field described above is a special case where the winding rate has no $z$-dependence (see Equation~(\ref{alphawind})) and, therefore, the radial average of the winding rate computed over a fixed averaging interval is independent of $z$, but this is not expected in general for fields that vary in the vertical direction.

From a different perspective, if the magnetic field varies in the vertical direction, the winding rate for a particular field line over a range of heights may be more important than the winding rate at a given radial location.
For an axisymmetric magnetic-flux tube with a field that varies in the $z$-direction, the radial locations of the surfaces of constant $\Psi$ generally vary in the $z$-direction (away from the axis), see, \myeg, Equation~(\ref{alphaeqn}).
Therefore, the winding rate for a particular field line at one height does not generally correspond to the winding rate of the same field line at some other height (away from the axis), see Figure~\ref{qlff}(b).
Again, from this perspective if the observed magnetic field varies in the vertical direction, the winding-rate estimates retrieved by fitting methods away from the flux-tube axis are generally expected to be valid only for the plane where the magnetic field is measured.

As mentioned above, the radial component of the magnetic field (see Equation~(\ref{alphabr})) is non-zero (for $m \neq 0$).
This suggests that the fitting model is not strictly appropriate for synthetic data constructed with this magnetic field.
However, the radial component is single-valued for a given value of $r$ at fixed height, which would indicate to the observer that the magnetic field is axisymmetric (as assumed by the fitting model).
The non-zero radial component of the magnetic field does not directly affect the winding-rate estimates.
This is because the winding rate at a given position is defined as the angle that a field line rotates about the flux-tube axis per unit length along the axis and, thus, we are fitting for the ratio $\wrat = B_\theta / B_z$, which does not involve $B_r$.
If a non-zero radial component of the magnetic field is detected in the measurements this indicates that the magnetic field may be varying with height and, therefore, the caveats discussed above apply (provided that the observed flux tube is axisymmetric with an axis that is straight and vertical).

\section{Testing the Fitting Methods with Twisted Magnetic Fields that Vary in the Azimuthal Direction}
\label{sec_ddthne0}

To show how the fitting methods perform when the observed flux tube has a magnetic field, and corresponding winding rate, that varies in the azimuthal direction we consider a magnetic field with:

\begin{equation}
B_{\theta} (r) = \qobsamp r B_0 \exp \left[ - 2 (r/ R)^2 \right] \, ,
\label{vartbteqn}
\end{equation}

\begin{equation}
B_z \left( r, \theta \right) = B_0 \exp \left[ - (r/ R)^2 \right] + \bvart  \left[ \left( r/ R \right)^2 - \left( r/R \right)^3 \right] \sin \left( \nvart \theta \right) \, ,
\label{vartbzeqn}
\end{equation}

\noindent
where 
$\qobsamp$ is a winding rate,
$B_0$ and $\bvart$ are magnetic-field strengths,
$R$ is the radius of the flux tube,
and
$\nvart$ is an integer.
At $r=0$ the magnetic field and corresponding current density are both parallel to the $z$-direction, and the Lorentz force is zero.
Away from $r=0$, the azimuthal component of this magnetic field [$B_{\theta}$] varies only in the radial direction, whereas the vertical component [$B_z$] can vary in both the radial and azimuthal directions (for $\bvart \neq 0$ and $\nvart \neq 0$).
This magnetic field is useful for testing the fitting methods because it has some features that are consistent with the fitting model (\myeg{} the flux-tube axis is straight and vertical, and the magnetic field does not vary in the vertical direction), which allows us to determine the influence of the variation of the magnetic field in both the azimuthal and radial directions on the winding-rate estimates retrieved with  fitting methods.

For $\bvart=0$ or $\nvart = 0$, the magnetic field described by Equations~(\ref{vartbteqn}) and (\ref{vartbzeqn}) is axisymmetric and the winding rate is $\wind (r) = \qobsamp \exp ( - (r/ R)^2 )$, which has the same radial dependence as the synthetic data used in Table~\ref{toytab}.
On the other hand, for $\bvart \neq 0$ and $\nvart \neq 0$, the winding rate of a magnetic-field line at the position $(r, \theta)$ about the flux-tube axis, per unit length along the axis, 

\begin{equation}
\wind
= \frac{B_\theta }{  r B_z }
= \qobsamp \exp \left[ - (r/ R)^2 \right] \left\{ 1 + (\bvart/B_0) \exp \left[ (r/ R)^2 \right] (r/ R)^2 \left[ 1 - \left( r/R \right) \right] \sin \left( \nvart \theta \right) \right\}^{-1} \, ,
\label{vartqeqn}
\end{equation}

\noindent
varies in both the $r$- and $\theta$-directions.
In either case, the winding rate at the flux-tube axis ($r=0$) is $\wind_{0,{\rm obs}} = \qobsamp$.

To generate synthetic data, we sample the magnetic field described by Equations~(\ref{vartbteqn}) and (\ref{vartbzeqn}) at a set of discrete locations on an $x$\,--\,$y$-plane, with a grid spacing of 100~km in both the $x$- and $y$-directions.
We set $B_0=3.5$~kG, $R=20$~Mm, and $\nvart=25$; these parameter values are chosen to be roughly consistent with a typical sunspot where azimuthal variations in the magnetic field are expected.
We use all points with $r_i \le R$ for the fitting, we calculate the radial average of the winding rate for the interval $[ 0 , R ]$,  and we choose the various parameter values for the synthetic data so that $B_z > 0$ within the flux tube.
This magnetic field may not closely resemble the types of fields expected in sunspots, but we emphasise that the point of this exercise is to quantify how azimuthal variations in this field affect the estimates for the on-axis and  average winding rates retrieved by fitting methods that use axisymmetric fitting models.

\begin{figure}[ht]
\hfil\includegraphics[width=0.6\textwidth]{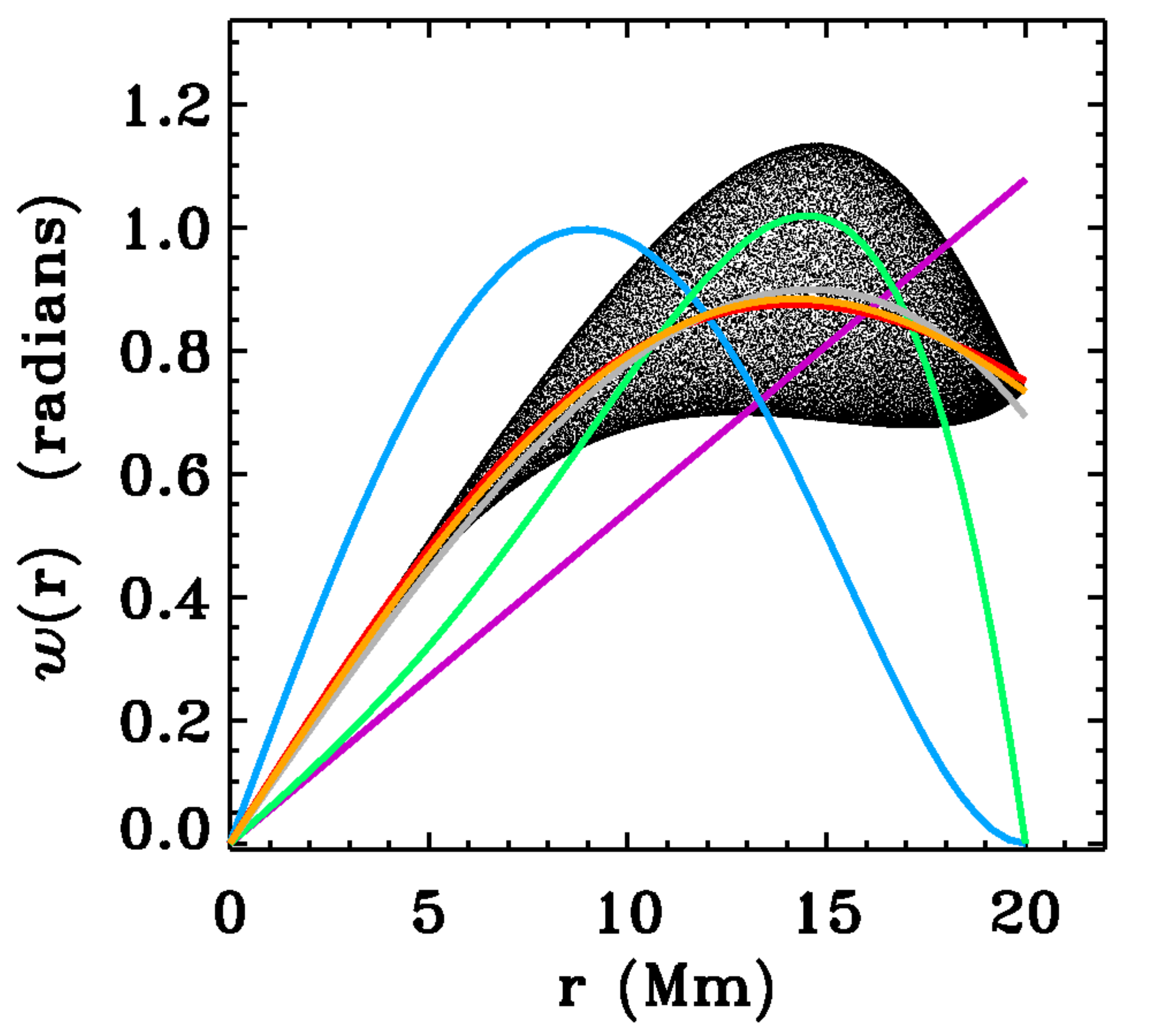}\hfil
\caption{The ratio, $\wrat = B_\theta / B_z$, as a function of $r$.
The black points correspond to the ratio $\wrat_{{\rm obs},i}$ derived from the synthetic measurements of a non-axisymmetric magnetic field (see Equations~(\ref{vartbteqn}) and (\ref{vartbzeqn})) with $\qobsamp=0.1$~radian~Mm$^{-1}$, $\bvart / B_0 = 1$, $R=20$~Mm, and $\nvart=25$.
The purple curve is the best-fit case for a fitting model with $\qmod(r) = \qmodamp$,
the blue curve is for $\qmod(r) = \qmodamp ( (1-(r/R)^2)^2 )$,
the green curve is for $\qmod(r) = \qmodamp (1 + 2 (r/R)^2 - 3 (r/R)^4 )$,
the red curve is for $\qmod(r) = \qmodamp \exp ( -(r/R)^2 )$,
the gray curve is for $\qmod(r) = \qbamp_1  + \qbamp_2 (r/R)^2$,
and the orange curve is for $\qmod(r) = \qbamp_1  + \qbamp_2 (r/R)^2 + \qbamp_3 (r/R)^4$, where $R$ is the radius of the fitting model (here $R=20$~Mm).
}
\label{vartplot}
\end{figure}

\begin{figure}[ht]
\includegraphics[width=0.49\textwidth]{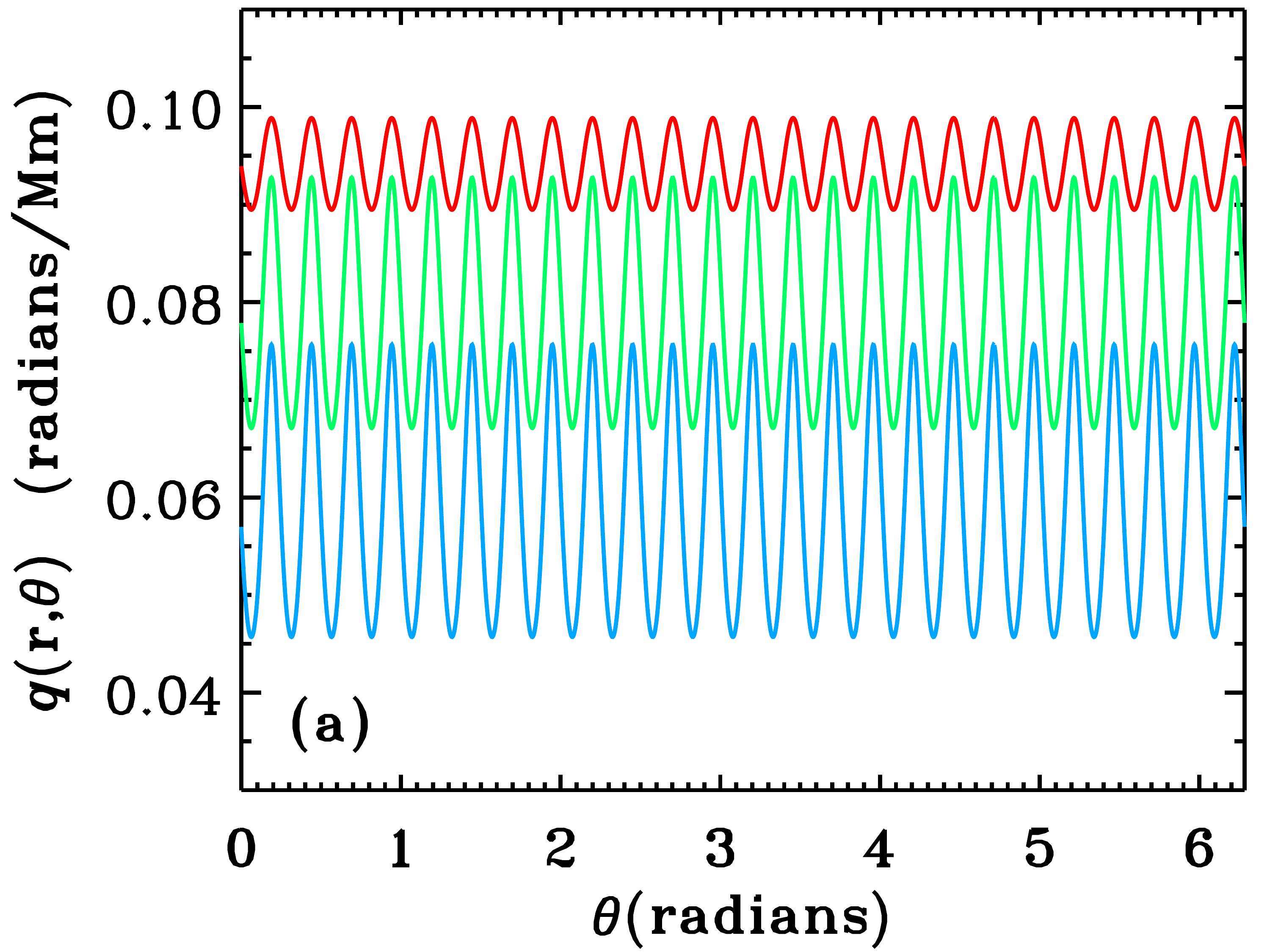}\hfil\includegraphics[width=0.49\textwidth]{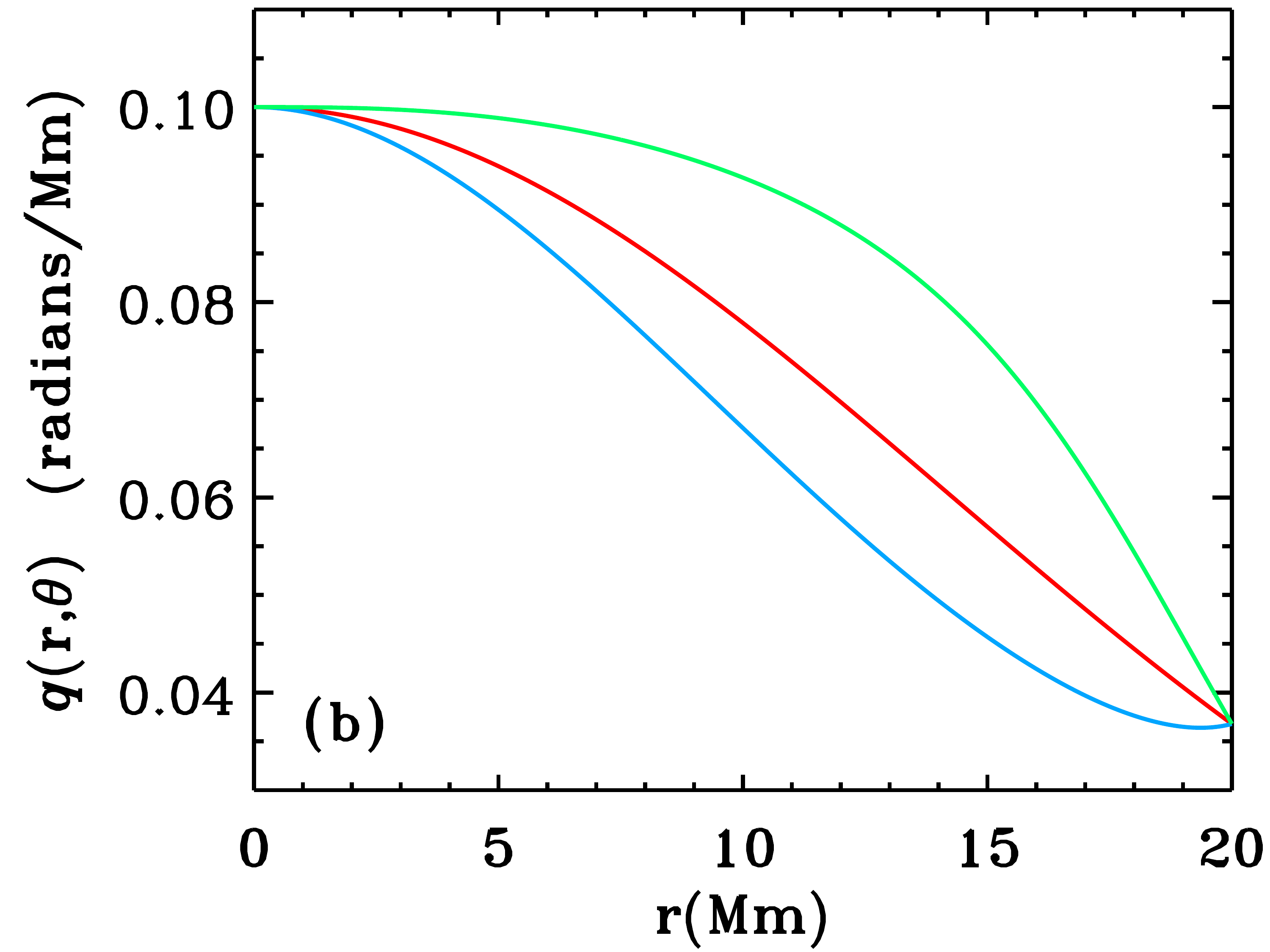}
\caption{(a) Winding rate [$\wind (r, \theta)$] as a function of $\theta$ at fixed $r$ for the non-axisymmetric magnetic field described in the text (see Equation~(\ref{vartqeqn})), with $B_0=3.5$~kG, $R=20$~Mm, $\bvart/B_0=1$, $\nvart=25$, and $\qobsamp =0.1$~radian~Mm$^{-1}$. 
The red, green, blue curves correspond to $r=$5, 10, 15~Mm, respectively.
(b) Winding rate [$\wind (r, \theta)$] as a function of $r$ at fixed $\theta$ for the same flux tube as (a).
The red, blue, green curves correspond to $\theta=$0 , $\pi / 50$ , $3 \pi / 50$~radians, respectively.
}
\label{vartqplot}
\end{figure}

Because the magnetic field and the winding rate vary in the azimuthal direction, the ratio $\wrat_{{\rm obs},i}$ derived from the synthetic measurements is not single-valued for a fixed value of $r$ (see Figures~\ref{vartplot} and \ref{vartqplot}).
The parameter $\bvart$ is related to the amplitude of the variation of the winding rate in the azimuthal direction (see Equation~(\ref{vartqeqn})), and we find that the scatter in the values of $\wrat_{{\rm obs},i}$ for a fixed value of $r$ tends to increase as $\bvart$ increases.
These features indicate that the fitting model is not strictly appropriate for synthetic data constructed with this magnetic field.
It may be possible to overcome this issue by extending the basis-function approximation approach of Section~\ref{sec_multi} to fit for both the azimuthal and radial dependence of the winding rate, but this is beyond the scope of the present investigation.
Instead, we apply the fitting methods discussed so far without modification (see, \myeg, Figure~\ref{vartplot}).
Because fitting methods that use axisymmetric fitting models cannot retrieve the azimuthal dependence of the winding rate, to test the efficacy of the fitting methods we use the azimuthally-averaged, radially-averaged winding rate of the observed magnetic field, along with the winding rate at the flux-tube axis.

\begin{table}
\caption{Results for tests of several fitting models applied to synthetic data constructed with the non-axisymmetric magnetic field described by Equations~(\ref{vartbteqn}) and (\ref{vartbzeqn}), with $B_0=3.5$~kG, $R=20$~Mm, and $\nvart=25$,  for three different values of $\bvart/B_0$.}
\label{varttab1}
\begin{tabular}{lccc}
\hline
Fitting model   & $ \wind_0^\ast /  \wind_{0,{\rm obs}} $ &  $ \wind_{\rm av}^\ast  / \wind_{\rm av,obs} $ & $\chi^2 / (R^2 \qobsamp^2)$\\[1mm]
\hline
\\[1mm]
$\bvart/B_0 =0$ & & & \\[1mm]
\\[1mm]
$\qmod (r) = \qmodamp$                                        & 0.529  & 0.708 & 0.00885 \\[1mm]
$\qmod (r) = \qmodamp (1-(r/R)^2)^2$                          & 1.709  & 1.220 & 0.0511 \\[1mm]
$\qmod (r) = \qmodamp  ( 1 + 2 (r/R)^2 - 3 (r/R)^4 )$         & 0.562  & 0.803 & 0.0115 \\[1mm]
$\qmod (r) = \qmodamp \exp (-(r/R)^2)$                        & 1.0  & 1.0 & 0.0 \\[1mm]
$\qmod (r) = \qbamp_1  + \qbamp_2 (r/R)^2$                    & 0.902  & 0.958 & 0.000138 \\[1mm]
$\qmod (r) = \qbamp_1  + \qbamp_2 (r/R)^2 + \qbamp_3 (r/R)^4$  & 0.988  & 0.996 & 9.586 $\times 10^{-7}$ \\[1mm]
\\[1mm]
\hline
\\[1mm]
$\bvart/B_0 =1$ & & & \\[1mm]
\\[1mm]
$\qmod (r) = \qmodamp$                                        & 0.539  & 0.714 & 0.0123 \\[1mm]
$\qmod (r) = \qmodamp (1-(r/R)^2)^2$                          & 1.741  & 1.231 & 0.0563 \\[1mm]
$\qmod (r) = \qmodamp  ( 1 + 2 (r/R)^2 - 3 (r/R)^4 )$         & 0.575  & 0.813 & 0.0142 \\[1mm]
$\qmod (r) = \qmodamp \exp (-(r/R)^2)$                        & 1.019  & 1.009 & 0.00304 \\[1mm]
$\qmod (r) = \qbamp_1  + \qbamp_2 (r/R)^2$                    & 0.923  & 0.969 & 0.00310 \\[1mm]
$\qmod (r) = \qbamp_1  + \qbamp_2 (r/R)^2 + \qbamp_3 (r/R)^4$  & 0.983  & 0.995 & 0.00303 \\[1mm]
\\[1mm]
\hline
\\[1mm]
$\bvart/B_0 =2$  & & & \\[1mm]
\\[1mm]
$\qmod (r) = \qmodamp$                                        & 0.574  & 0.736  & 0.0274 \\[1mm]
$\qmod (r) = \qmodamp (1-(r/R)^2)^2$                          & 1.852  & 1.265  & 0.0780 \\[1mm]
$\qmod (r) = \qmodamp  ( 1 + 2 (r/R)^2 - 3 (r/R)^4 )$         & 0.619  & 0.845  & 0.0265 \\[1mm]
$\qmod (r) = \qmodamp \exp (-(r/R)^2)$                        & 1.088  & 1.040 & 0.0166 \\[1mm]
$\qmod (r) = \qbamp_1  + \qbamp_2 (r/R)^2$                    & 0.997  & 1.006   & 0.0163 \\[1mm]
$\qmod (r) = \qbamp_1  + \qbamp_2 (r/R)^2 + \qbamp_3 (r/R)^4$  & 0.960  & 0.990  & 0.0163 \\[1mm]
\hline
\end{tabular}
\end{table}

For synthetic data generated with this magnetic field, it can be shown that the best-fit on-axis and radially-averaged winding rates are directly proportional to the true on-axis and true azimuthally averaged, radially averaged winding rates, respectively.
In Table~\ref{varttab1} we show the ratio of the inferred and true winding rates at the flux-tube axis [$\wind_0^\ast / \wind_{0,{\rm obs}}$] along with the ratio of the inferred radially averaged winding rate to the true azimuthally averaged, radially averaged winding rate [$\wind_{\rm av}^\ast / \wind_{\rm av,{\rm obs}}$] for three different values of $\bvart/B_0$.
We include the axisymmetric case (with $\bvart=0$) for reference and because the magnetic field for these tests is sampled  differently to the case shown in Table~\ref{toytab}.
We find that the results of these tests are not strongly sensitive to the value of $\nvart$ (although this depends on how the magnetic field is sampled).
On the other hand, it is clear from Table~\ref{varttab1} that the results are sensitive to the value of $\bvart/B_0$.

For the case with $\bvart/B_0=1$, despite the presence of moderate azimuthal variations in the magnetic field and winding rate, we find the same general trend as in the cases that use axisymmetric synthetic data [$\bvart=0$].
For example, we find that both the on-axis and average winding rates retrieved by the best-fit polynomial basis-function approximations are generally more accurate (and the $\chi^2$ values are smaller) than those retrieved with fitting models with a fixed radial variation for the winding rate.
The exception is for the fitting model with $\qmod (r) = \qmodamp \exp (-(r/R)^2)$, which retrieves accurate winding-rate estimates since it matches the winding-rate profile for the axisymmetric version of these  data.
We also find  that the various fitting methods that make fixed assumptions about the radial variation of the winding rate can produce different estimates for the winding rates given the same synthetic data.

For a given fitting model, the discrepancy between the inferred and true winding rates does not necessarily increase as the value of $\bvart/B_0$ increases (as one may expect), although the magnitude of the $\chi^2$ values does generally increase (see Table~\ref{varttab1}).
For the best-fit polynomial basis-function approximations applied to synthetic data with $\bvart/B_0=2$, we find that the discrepancy between the inferred and true winding rates increases as $n_b$ increases, contrary to the trend found at lower values of $\bvart/B_0$.
Nevertheless, we generally find for the best-fit polynomial basis-function approximations that the discrepancy between the inferred and true winding rates is less than  10\,\% for $\bvart/B_0=2$ (see Table~\ref{varttab1}).
Those discrepancies are quite small considering that the winding rate varies approximately by a factor of three from its minimum to maximum value for some values of $r$ for $\bvart/B_0=2$; for the case with $\bvart/B_0=1$ the winding rate can vary by a factor of roughly two (see Figure~\ref{vartqplot}(a)). 
We find similar results in other experiments  with different functional forms for both the azimuthal and radial dependence of the magnetic field as those used in Equations~(\ref{vartbteqn}) and (\ref{vartbzeqn}) (results not shown).

\section{Testing the Role of the Location of the Fitting-Model Axis}
\label{sec_axisloc}

Some methods that are commonly used to estimate the location of the axis of a magnetic-flux tube are:
the peak of $| \B |$, 
the peak of $| B_z |$, 
the flux-weighted centroid, 
and
the peak of the force-free parameter $\alpha_z$
(\myeg{} \opencite{lea03}; \opencite{kink}; \opencite{Nandyetal2008}).
As discussed by \inlinecite{kink} some of these methods do not consistently recover the correct location of the flux-tube axis; we have confirmed this finding using synthetic data constructed with a model for a toroidal magnetic-flux loop (\myeg{} \opencite{fan03}, \citeyear{fan04}, and see Section~\ref{sec_tor}), but do not show the results for the sake of brevity.
To demonstrate how the winding-rate estimates  are affected by an error in the location of the fitting-model axis (in the absence of other sources of error), we generate synthetic data using a magnetic field with

\begin{equation}
B_\theta \left( r \right) = \qobsamp r B_0 \exp \left[ - 2 (r / R)^2 \right]   \quad \mbox{and} \quad B_z \left( r \right) = B_0 \exp \left[ - (r / R)^2 \right] \, , 
\label{axisloceqn}
\end{equation}

\noindent
where
$\qobsamp$ is a  winding rate,
$B_0$ is a magnetic-field strength,
and
$R$ is the radius of the flux tube.
To generate synthetic data we set $\qobsamp=0.1$~radians~Mm$^{-1}$, $B_0=3.5$~kG, and $R=20$~Mm, and sample the magnetic field at a set of discrete locations on an $x$\,--\,$y$-plane, with a grid spacing of 100~km in both the $x$- and $y$-directions.
We introduce an error into the location of the flux-tube axis by computing  synthetic data with the axis of the flux tube located at $x=4$~Mm and $y=0$.
We apply the various fitting methods  without modification assuming that the fitting-model axis is located at $x=0$ and $y=0$.
To estimate the winding rates we use only observation points that lie within 20~Mm of the fitting-model axis and have $B_z > 0$. 
We calculate the average winding rate over the interval used for the fitting (\myie{} $0 \le r \le \max (r_i)$).

\begin{figure}[ht]
\hfil\includegraphics[width=0.6\textwidth]{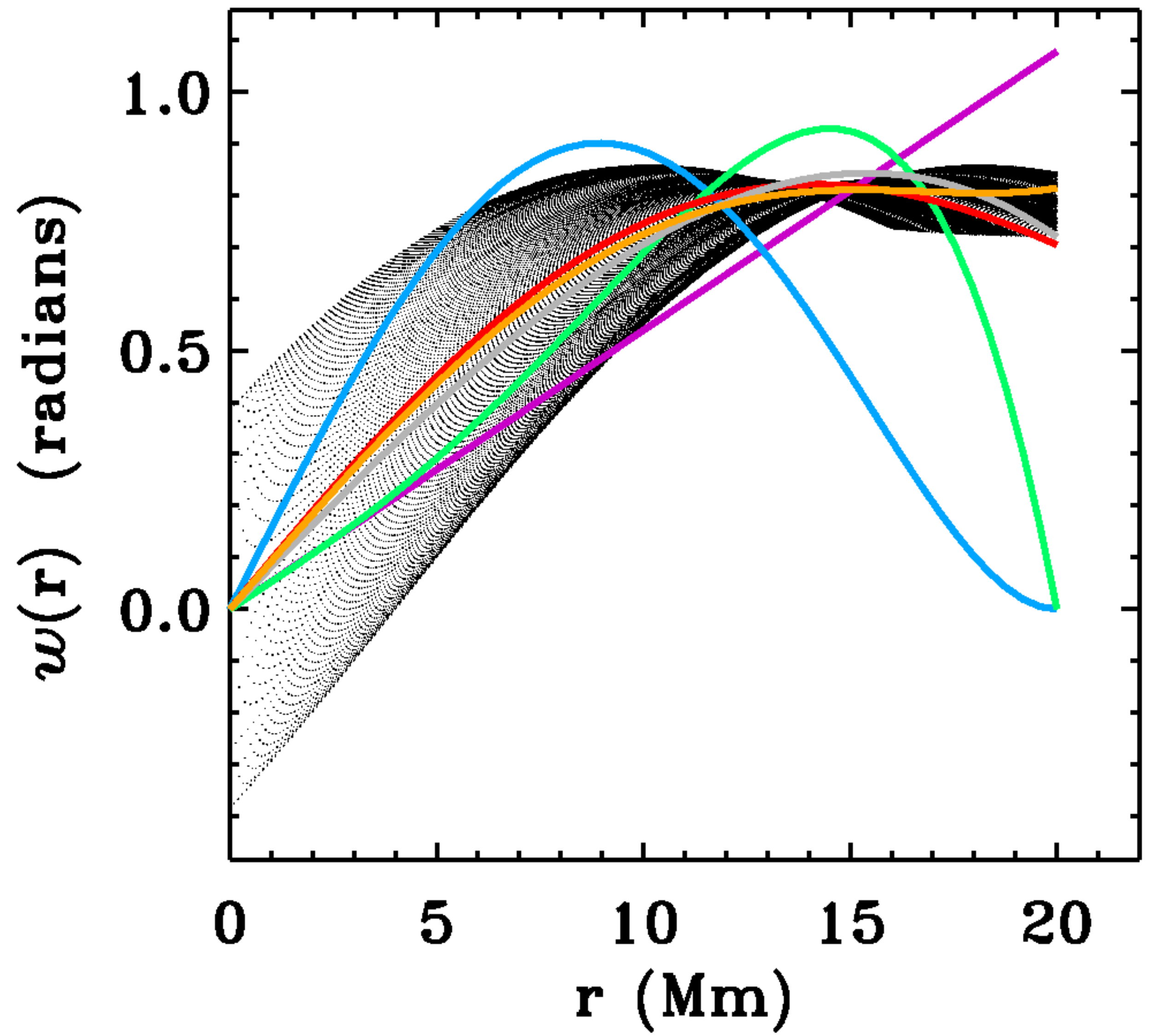}\hfil
\caption{
The ratio $\wrat = B_\theta / B_z$ as a function of $r$ (the radial distance from the fitting-model axis) for a test where an error is  introduced into the location of the fitting-model axis, see Appendix~\ref{sec_axisloc} for details.
The black points correspond to the ratio $\wrat_{{\rm obs},i}$ derived from the synthetic measurements.
The purple curve is the best-fit case for a fitting model with $\qmod(r) = \qmodamp$,
the blue curve is for $\qmod(r) = \qmodamp ( (1-(r/R)^2)^2 )$,
the green curve is for $\qmod(r) = \qmodamp (1 + 2 (r/R)^2 - 3 (r/R)^4 )$,
the red curve is for   $\qmod(r) = \qmodamp  \exp ( -(r/R)^2)$,
the gray curve is for $\qmod(r) = \qbamp_1  + \qbamp_2 (r/R)^2$,
and the orange curve is for $\qmod(r) = \qbamp_1  + \qbamp_2 (r/R)^2 + \qbamp_3 (r/R)^4$, where $R$ is the radius of the fitting model (here $R=20$~Mm).
}
\label{axisloc}
\end{figure}

When there is an error in the location of the fitting-model axis we find that $\wrat$ is not single-valued for a fixed value of $r$ (see Figure~\ref{axisloc}), which indicates that the fitting model is not strictly appropriate for these synthetic data.
Nevertheless,  it can be shown that the best-fit on-axis and average winding rates are directly proportional to the respective true values for this test.
In Table~\ref{axisloctab} we provide the ratios $\wind_0^\ast / \wind_{0,{\rm obs}}$ and $\wind_{\rm av}^\ast / \wind_{\rm av,obs}$, along with the corresponding value for $\chi^2 / (R^2 \qobsamp^2)$ for the various best-fit models shown in Figure~\ref{axisloc}.
The magnetic field described by Equations~(\ref{axisloceqn}) is the same as for the axisymmetric case (with $\bvart=0$) used in Appendix~\ref{sec_ddthne0}; hence, for reference,  the results in Table~\ref{varttab1} for the case with $\bvart=0$ are those that would be retrieved if the fitting-model axis were correctly located for this synthetic data.
Evidently, the accuracy of the winding-rate estimates retrieved by the fitting methods is affected by an error in the location of the fitting-model axis.
However, for the better-performing cases (such as the polynomial basis-function approximations with $n_b=3$ and the case with the correct winding-rate profile) the magnitude of the relative discrepancy caused by the error in the location of the fitting-model axis is less than 10\,\%. 
We find that the magnitude of the discrepancy in the inferred winding rates generally gets progressively larger as the error introduced into the axis location is increased.
In experiments, using synthetic data generated with winding-rate profiles and magnetic fields different from Equation~(\ref{axisloceqn}), we find that an error in the location of the fitting-model axis generally produces an error in the winding-rate estimates retrieved by the fitting methods of comparable magnitude to the case discussed here.

\begin{table}
\caption{Summary of results for a test where an error is  introduced into the location of the fitting-model axis, see Appendix~\ref{sec_axisloc} for details.
}
\label{axisloctab}
\begin{tabular}{llll}
\hline
Fitting model     & $\wind_0^\ast /  \wind_{0,{\rm obs}}$ &  $\wind_{\rm av}^\ast  / \wind_{\rm av,obs} $ & $\chi^2 / (R^2 \qobsamp^2)$ \\[1mm]
\hline
$\qmod (r) = \qmodamp$                                        & 0.539 & 0.722  & 0.00841 \\[1mm]
$\qmod (r) = \qmodamp (1-(r/R)^2)^2$                          & 1.573 & 1.124  & 0.0445 \\[1mm]
$\qmod (r) = \qmodamp  ( 1 + 2 (r/R)^2 - 3 (r/R)^4 )$         & 0.524 & 0.748  & 0.0121 \\[1mm]
$\qmod (r) = \qmodamp  \exp ( -(r/R)^2)$                      & 0.957 & 0.957  & 0.00323 \\[1mm]
$\qmod (r) = \qbamp_1  + \qbamp_2 (r/R)^2$                    & 0.820 & 0.893  & 0.00323 \\[1mm]
$\qmod (r) = \qbamp_1  + \qbamp_2 (r/R)^2 + \qbamp_3 (r/R)^4$  & 0.926 & 0.940  & 0.00301 \\[1mm]
\hline
\end{tabular}
\end{table}


\end{article}
\end{document}